\documentclass[fleqn,usenatbib]{mnras}

\usepackage{newtxtext,newtxmath}

\usepackage[T1]{fontenc}

\DeclareRobustCommand{\VAN}[3]{#2}
\let\VANthebibliography\thebibliography
\def\thebibliography{\DeclareRobustCommand{\VAN}[3]{##3}\VANthebibliography}

\usepackage{graphicx}	
\usepackage{amsmath} 
 
\usepackage{amssymb}	

\usepackage{siunitx}
\usepackage{color,colortbl}
\usepackage{multirow}
\usepackage{multicol}
\usepackage{hyperref}
\usepackage{longtable}
\usepackage[dvipsnames]{xcolor}

\DeclareSIUnit \parsec {pc}

\DeclareMathOperator\erf{erf}

\title[DEVILS: Identifying AGN through SED Fitting]{Deep Extragalactic VIsible Legacy Survey (DEVILS): Identification of AGN through SED Fitting and the Evolution of the Bolometric AGN Luminosity Function}

\author[J. E. Thorne et al.]{
Jessica E. Thorne,$^{1,2}$\thanks{E-mail: jessica.thorne@icrar.org}
Aaron S. G. Robotham,$^{1,2}$
Luke J. M. Davies,$^{1}$
Sabine Bellstedt,$^{1}$\newauthor
Michael J. I. Brown,$^{3}$
Scott M. Croom, $^{4,2}$
Ivan Delvecchio,$^{5}$
Brent Groves,$^{1}$
Matt J. Jarvis,$^{6,7}$\newauthor
Stanislav S. Shabala,$^{8,2}$
Nick Seymour,$^{9}$
Imogen H. Whittam,$^{6,7}$
Matias Bravo,$^{1}$
Robin H. W. Cook,$^{1}$\newauthor
Simon P. Driver,$^{1}$
Benne Holwerda,$^{10}$
Steven Phillipps,$^{11}$
Malgorzata Siudek$^{12,13}$
\\
$^{1}$ ICRAR, The University of Western Australia, 35 Stirling Highway, Crawley, WA 6009, Australia\\
$^{2}$ ARC Centre of Excellence for All Sky Astrophysics in 3 Dimensions (ASTRO 3D)\\
$^{3}$ School of Physics and Astronomy, Monash University, VIC 3800, Australia\\
$^{4}$ Sydney Institute for Astronomy, School of Physics, A28, The University of
Sydney, Sydney, NSW 2006, Australia\\
$^{5}$ INAF - Osservatorio Astronomico di Brera, Via Brera 28, 20121 Milano, Italy\\
$^{6}$ Astrophysics, University of Oxford, Denys Wilkinson Building, Keble Road, Oxford, OX1 3RH, UK\\
$^{7}$ Department of Physics and Astronomy, University of the Western Cape, Robert Sobukwe Road, Bellville 7535, South Africa\\
$^{8}$ School of Natural Sciences, Private Bag 37, University of Tasmania, Hobart, TAS 7001, Australia\\
$^{9}$ ICRAR, Curtin University, Bentley, WA 6102, Australia\\
$^{10}$ Department of Physics and Astronomy, 102 Natural Science Building, University of Louisville, Louisville KY 40292, USA\\
$^{11}$ Astrophysics Group, School of Physics, University of Bristol, Bristol BS8 1TL, UK \\
$^{12}$ Institut de F\'{\i}sica d'Altes Energies (IFAE), The Barcelona Institute of Science and Technology, 08193 Bellaterra (Barcelona), Spain \\
$^{13}$ National Centre for Nuclear Research, ul. Pasteura 7, 02-093, Warsaw, Poland
}

\date{Accepted XXX. Received YYY; in original form ZZZ}

\pubyear{2021}

\begin{document}
\label{firstpage}
\pagerange{\pageref{firstpage}--\pageref{lastpage}}
\maketitle

\begin{abstract}
Active galactic nuclei (AGN) are typically identified through radio, mid-infrared, or X-ray emission or through the presence of broad and/or narrow emission lines.
AGN can also leave an imprint on a galaxy’s spectral energy distribution (SED) through the re-processing of photons by the dusty torus.
Using the SED fitting code \textsc{ProSpect} with an incorporated AGN component, we fit the far ultraviolet to far-infrared SEDs of $\sim$494,000 galaxies in the D10-COSMOS field and $\sim$230,000 galaxies from the GAMA survey.
By combining an AGN component with a flexible star formation and metallicity implementation, we obtain estimates for the AGN luminosities, stellar masses, star formation histories, and metallicity histories for each of our galaxies. 
We find that \textsc{ProSpect} can identify AGN components in 91 per cent of galaxies pre-selected as containing AGN through narrow-emission line ratios and the presence of broad lines. 
Our \textsc{ProSpect}-derived AGN luminosities show close agreement with luminosities derived for X-ray selected AGN using both the X-ray flux and previous SED fitting results. 
We show that incorporating the flexibility of an AGN component when fitting the SEDs of galaxies with no AGN has no significant impact on the derived galaxy properties. 
However, in order to obtain accurate estimates of the stellar properties of AGN host galaxies, it is crucial to include an AGN component in the SED fitting process.
We use our derived AGN luminosities to map the evolution of the AGN luminosity function for $0<z<2$ and find good agreement with previous measurements and predictions from theoretical models. 
\end{abstract}





\section{Introduction}\label{sec:Intro}
It is widely accepted that every massive galaxy hosts a supermassive black hole (SMBH; $>10^6M_\odot$) in its centre \citep{MagorrianDemographyMassiveDark1998,FilippenkoLowMassCentralBlack2003,KormendySecularEvolutionFormation2004,BarthPOX52Dwarf2004,GreeneActiveGalacticNuclei2004,GreeneNewSampleLowMass2007,GreeneBlackHolesPseudobulges2008}. 
While some of these appear to be non-accreting, such as the SMBH at the centre of our Galaxy, others are accreting and termed Active Galactic Nuclei (AGN). 
AGN have been detected across the entire electromagnetic spectrum spanning the gamma ray to radio wavelengths and in their most active phases can outshine their host galaxy \citep{PadovaniActivegalacticnuclei2017}.

Since there are several mechanisms involved in AGN emission, they emit across the electromagnetic spectrum and can therefore be detected through many observational techniques, such as through broad and/or narrow optical emission lines (e.g. \citealt{BaldwinClassificationparametersemissionline1981,KewleyTheoreticalModelingStarburst2001,MeyerQuasarscompletespectroscopic2001,Kauffmannhostgalaxiesactive2003}), X-ray emission (e.g. \citealt{BrandtCosmicXraysurveys2015}), mid-infrared emission (e.g. \citealt{LacyObscuredUnobscuredActive2004,SternMidInfraredSelectionActive2005,JarrettSPITZERWISESURVEY2011,SternMIDINFRAREDSELECTIONACTIVE2012,AssefMidInfraredSelectionActive2013,LacySpitzerMidinfraredAGN2015}), and radio emission (e.g. \citealt{HeckmanCoevolutionGalaxiesSupermassive2014,Padovanifaintradiosky2016,TadhunterRadioAGNlocal2016}).
These processes are to some extent uncoupled, making unambiguous classification of AGN difficult. 
The detection and classification of AGN also depends on the environment and geometry of the system as different lines of sight result in different observed emission processes. 
While we adopt the generic AGN terminology here, historically various geometric configurations and detection techniques have led to a plethora of disparate naming conventions \citep{PadovaniActivegalacticnuclei2017}.

One of the newest forms of AGN identification is through simultaneously fitting the galaxy and possible AGN contribution to the far ultraviolet (FUV) to far infrared (FIR) spectral energy distribution (SED; see \citealt{PouliasisobscuredAGNpopulation2020,MountrichasXrayfluxSED2021}).
Although the FUV-FIR SED of galaxies is usually dominated by contributions from stellar emission and re-processing of stellar light by dust, in some cases AGN can have a significant impact on the shape of the SED in the mid-infrared (MIR; e.g. \citealt{PollettaChandraSpitzerUnveil2006,Brownspectralenergydistributions2019}).
The MIR emission from AGN originates from the re-processing of accretion disc photons by the dusty torus, which is generally at hotter temperatures than dust in the diffuse interstellar medium or in proximity to recent star formation \citep{LaorSpectroscopicconstraintsproperties1993,NenkovaDustEmissionActive2002}.
Simple MIR broad-band colour selection techniques have proven extremely useful in revealing the presence of an AGN \citep{LacyOpticalSpectroscopyXRay2007,JarrettSPITZERWISESURVEY2011,SternMIDINFRAREDSELECTIONACTIVE2012, DonleyIdentifyingLuminousActive2012, AssefMidInfraredSelectionActive2013,AssefWISEAGNCatalog2018}. 
However, broad-band MIR selection techniques are biased against low luminosity AGN if the stellar light is bright \citep{BarmbyMidInfraredPropertiesXRay2006,GeorgantopoulosSearchingmidIRobscured2008}. 
SED fitting can alleviate this problem by using a larger wavelength range and can disentangle AGN emission from the host galaxy emission. 

Two different modes of AGN FUV-FIR SED fitting codes exist, each designed for different purposes - those that are designed to fit the host galaxy with an added AGN model, and those designed to fit an AGN component with the bare necessities for the host galaxy. 
The first generally makes more assumptions about the emission and geometry of the AGN and include codes such as \textsc{Prospector} \citep{Prospector,JohnsonStellarPopulationInference2021} and \textsc{Cigale} \citep{CIGALE,BoquienCIGALEpythonCode2019,YangxcigalefittingAGN2020} as used in \cite{LejaHotDustPanchromatic2018} and \cite{CieslaConstrainingpropertiesAGN2015,PouliasisobscuredAGNpopulation2020} for example. 
The second type generally makes more simplifications about the star formation history, metallicity, and dust of the host galaxy and includes codes such as \textsc{AGNfitter} \citep{CalistroRiveraAGNfitterBayesianMCMC2016}. 

While many codes exist to disentangle the contribution of the AGN and host galaxy to the FUV-FIR SED, there is no single simple way of including an AGN in the SED fitting process. 
The power-law emission from an AGN accretion disc in the rest-frame UV-optical is often difficult to distinguish from the emission from actively star forming galaxies \citep{CardosoImpactAGNfeatureless2017}, and the re-processing of emission by the dusty torus can present in a similar manner to the re-processing by  dust in a galaxy's interstellar medium. 
The geometry of the system also needs to be considered as obscuration of the central engine by dust can have a significant impact on the shape of the SED.

As mentioned above, SED fitting has the potential to identify lower luminosity AGN than simple MIR broad-band colour selections due to the larger wavelength range used and because it can properly separate the star formation and AGN components. 
SED fitting techniques can also identify highly obscured AGN populations that are not identifiable in X-ray emission \citep{PouliasisobscuredAGNpopulation2020} but will be unable to detect extremely obscured low-luminosity objects.

A fundamental diagnostic for studying the evolution of AGN is the AGN luminosity function and its evolution with redshift. 
The luminosity function of AGN has been studied for decades in the rest-frame optical/UV \citep[e.g.][]{SchmidtSpaceDistributionLuminosity1968, SchmidtQuasarevolutionderived1983,KooSpectroscopicsurveyQSOs1988,Boyleevolutionopticallyselected1988,RichardsSloanDigitalSky2006,Croom2dFSDSSLRGQSO2009, RossSDSSIIIBaryonOscillation2013,McGreerQuasarLuminosityFunction2013}, soft X-ray \citep[e.g.][]{MaccacaropropertiesXrayselectedactive1991,BoyledeepROSATsurvey1993,JonesXrayQSOevolution1997,PageevolutionQSOsderived1997,MiyajiSoftXrayAGN2000,HasingerLuminositydependentevolutionsoft2005}, hard X-ray  \citep[e.g.][]{UedaCosmologicalEvolutionHard2003,LaFrancaHELLAS2XMMSurveyVII2005,BargerNumberDensityIntermediate2005,SilvermanLuminosityFunctionXRayselected2008,EbreroXMMNewtonserendipitoussurvey2009,YenchoOPTXProjectII2009,AirdevolutionhardXray2010,AirdNuSTARExtragalacticSurvey2015}, IR \citep[e.g.][]{BrownInfraredLuminosityFunction2006,MatuteActivegalacticnuclei2006,AssefMidIRXraySelectedQSO2011,LacySpitzerMidinfraredAGN2015} and radio \citep[e.g.][]{Rigbyluminositydependenthighredshiftturnover2011,SmolcicVLACOSMOSGHzLarge2017}.
These studies have conclusively shown that the observed AGN luminosity function has a strong redshift evolution in both normalisation (number density) but also in the slope. 
The number density of low luminosity AGN peaks at a lower redshift than that of bright AGN indicating the `cosmic downsizing' of AGN \citep{HasingerLuminositydependentevolutionsoft2005,BargerNumberDensityIntermediate2005,BabicObservationalEvidenceActive2012}. 
AGN feedback, which can limit the supply of gas for accretion, may be responsible for this phenomenon. 
Recent studies have pushed the study of the AGN luminosity function to $z > 7$ \citep{Mortlockluminousquasarredshift2011,Bowlergalaxyluminosityfunction2015, Banados800millionsolarmassblackhole2018,WangDiscoveryLuminousBroad2018,Bowlerlackevolutionvery2020} revealing the early growth of SMBHs.

Interpreting the various AGN luminosity functions is complicated by the fact that observations in a single band are always subject to selection effects and host galaxy contamination. 
Although AGN are intrinsically luminous across the UV-optical, dust extinction along certain viewing angles can make AGN difficult to detect. 
Isolating emission from the AGN in the UV-optical can also be difficult due to contamination from the host galaxy's stellar light.
Even in the X-ray, which is much less impacted by dust than the UV-optical, Compton-thick AGN, which can account for 20-50 per cent of the total AGN population \citep{BurlonThreeyearSwiftBATSurvey2011,RicciComptonthickAccretionLocal2015}, are still severely obscured and current observations remain largely incomplete. 
In the MIR and FIR, observations can be contaminated by the dust emission in the host galaxy, again limiting the effectiveness of AGN identification and also our ability to measure accurate AGN luminosities. 
Previous studies have combatted this by combining measurements of the AGN luminosity function in various wavelengths into a single `bolometric luminosity function' \citep{HopkinsObservationalDeterminationBolometric2007,ShankarSELFCONSISTENTMODELSAGN2009, Shenbolometricquasarluminosity2020}.
This allows for a higher level of completeness across selection techniques but relies on various bolometric conversions to combine measurements.

By identifying AGN through multiwavelength SED fitting, the emission from an AGN and the host galaxy across the UV-FIR can be simultaneously modelled, allowing for constraint on the AGN emission in both the UV and MIR. 
Simultaneous modelling of the AGN and host galaxy emission allows for extraction of AGN luminosities that are less likely to be contaminated by the host galaxy. 
This technique is, however, still limited to AGN with emission in the FUV-MIR distinguishable from that of the host galaxy.

In this work, we apply a new version of the SED fitting code \textsc{ProSpect} \citep{RobothamProSpectgeneratingspectral2020} with an incorporated AGN model to a sample of 494,000 galaxies in the DEVILS D10-COSMOS field \citep{DaviesDeepExtragalacticVIsible2018} spanning $0 < z <9$ to identify AGN through their SEDs. 
We also fit the SEDs of 230,000 galaxies from the GAMA survey spanning $0 < z < 4$, but with 95 per cent of sources with $z < 0.5$.
This work will describe the SED fitting and AGN selection techniques used to isolate a sample of AGN.
We will also provide validation of the technique and demonstrate the impact of an AGN component on derived galaxy properties. 
In a follow-up paper we will use these SED selected AGN to investigate the properties of AGN host galaxies (Thorne et al. in prep).
The structure of this paper is as follows.
After describing the DEVILS project and related data sets in Section~\ref{sec:Data}, we describe the SED fitting method and AGN model used in this work in Section~\ref{sec:SED}.
We compare the AGN identified and quantified by \textsc{ProSpect} to other techniques in Sections~\ref{sec:comparisons} and \ref{sec:AGNQuantification} respectively. 
We investigate the impact of an AGN component on the derived galaxy properties in Section~\ref{sec:EffectOnHostGalaxies}. 
Using our derived AGN luminosities we derive the AGN luminosity function and evolution of the AGN luminosity density as further validation of our derived AGN luminosities in Section~\ref{sec:AGNLF}.
We summarise our results in Section~\ref{sec:conclusion}.
Throughout this work we use a \cite{ChabrierGalacticStellarSubstellar2003} IMF and all magnitudes are quoted in the AB system unless stated. 
We adopt the \cite{PlanckCollaborationPlanck2015results2016} cosmology with $H_0 = 67.8 \, \si{\kilo \meter \per \second \per \mega \parsec}$, $\Omega_{M} = 0.308$ and $\Omega_\Lambda = 0.692$. 

\section{Data} \label{sec:Data}

For this work we use the Deep Extragalactic VIsible Legacy Survey (DEVILS; \citealt{DaviesDeepExtragalacticVIsible2018}) and the Galaxy and Mass Assembly (GAMA) Survey \citep{DriverGalaxyMassAssembly2011,LiskeGalaxyMassAssembly2015}.
In the following subsections we describe the DEVILS and GAMA data sets.

\subsection{Deep Extragalactic VIsible Legacy Survey}
DEVILS is an on-going optical spectroscopic redshift survey using the Anglo-Australian Telescope specifically designed to have high spectroscopic completeness over a large redshift range ($0.3 < z < 1$).
DEVILS targets three well-studied extragalactic fields: COSMOS (D10,1.5 deg$^2$),  ECDFS (D02, 3 deg$^2$), and XMM-LSS (D03, 1.5 deg$^2$) covering a total of 6 deg$^2$. 
These three fields were selected due to the wealth of existing multiwavelength data covering the X-ray to radio regimes. 
DEVILS will build a spectroscopic sample of $\sim 60,000$ galaxies down to $Y_{\text{mag}} < 21.2$ to a high completeness ($> 85$ per cent), allowing for robust parameterization of group and pair environments in the distant universe. For a full description of the survey science goals, survey design, target selection and spectroscopic observations see \cite{DaviesDeepExtragalacticVIsible2018}. 

In this work, as in \cite{ThorneDeepExtragalacticVIsible2021}, we use the spectroscopic and photometric data from the D10-COSMOS field. 
The D10-COSMOS field was prioritised for early science due to the large number of existing spectroscopic redshifts from previous surveys (i.e. zCOSMOS; \citealt{LillyzCOSMOS10kBrightSpectroscopic2009}) and as such currently has the highest spectroscopic completeness.
We use the new DEVILS photometry catalogue as described in depth by \cite{DaviesDeepExtragalacticVIsible2021} which uses the \textsc{ProFound} source extraction code \citep{RobothamProFoundSourceExtraction2018} to detect sources and measure source photometry consistently in 22 bands spanning the FUV-FIR (1500\AA- 500 $\mu$m). 
For D10 we use imaging in the GALEX \textit{FUV NUV} \citep{ZamojskiDeepGALEXImaging2007}, CFHT \textit{u} \citep{CapakFirstReleaseCOSMOS2007}, Subaru HSC \textit{griz} \citep{AiharaSeconddatarelease2019}, VISTA 
\textit{YJHK$_{s}$} \citep{McCrackenUltraVISTAnewultradeep2012}, Spitzer \textit{IRAC1 IRAC2 IRAC3 IRAC4 MIPS24 MIPS70} \citep{SandersSCOSMOSSpitzerLegacy2007,LaigleCOSMOS2015CATALOGEXPLORING2016}, and Herschel \textit{P100 P160 S250 S350 S500} \citep{LutzPACSEvolutionaryProbe2011,OliverHerschelMultitieredExtragalactic2012} bands.
Photometry was extracted in two phases to account for large differences in resolution and depth between the FUV-NIR and MIR-FIR regimes.
Photometry for the \textit{FUV-IRAC4} bands was extracted using the segment mode in \textsc{ProFound} while the photometry for the \textit{MIPS24-S500} bands was extracted using the \texttt{FitMagPSF} mode in \textsc{ProFound}. 
Due to poor resolution and shallow imaging, FIR photometry was only extracted for optically selected objects with $Y<21.2$mag or that were detected in the MIPS24 imaging. 
If an object met the criteria for FIR photometry extraction it was passed through the \texttt{FitMagPSF} mode.
Objects that met the FIR photometry criteria but not detected in a FIR band have a flux measurement of zero and a flux error measured from the sky noise. 
If no attempt was made to measure FIR photometry for an object then it will have no value for the flux and flux error.
5 per cent of all optically-detected objects used in this met the FIR photometry criteria and therefore have FIR photometry measurements.

As described in \cite{ThorneDeepExtragalacticVIsible2021}, we use a compilation of spectroscopic, grism and photometric redshifts to allow for galaxy properties to be estimated for as many galaxies as possible. 
The various redshift sources are presented in table C1 of \cite{ThorneDeepExtragalacticVIsible2021} and include 3,394 spectroscopic redshifts measured as part of the DEVILS program.

As in \cite{ThorneDeepExtragalacticVIsible2021}, we remove all segments classed by \cite{DaviesDeepExtragalacticVIsible2021} as stars (\textit{starflag} column, 16\,158 segments), artefacts (\textit{artefactflag}, 11\,072 segments) or that are masked (\textit{mask}, 120\,863 segments). 
Each of these flags are described in detail by \cite{DaviesDeepExtragalacticVIsible2021}, but briefly, stars and ambiguous objects are identified through size and colour, with cuts defined using the source type derived from the photometric redshift fitting code \textit{Le Phare} from the COSMOS2015 catalogue \citep{LaigleCOSMOS2015CATALOGEXPLORING2016}.\footnote{There are $\sim1,000$ objects (0.2 per cent) classed as stars due to small sizes but have galaxy colours which could potentially be AGN (QSOs). }
Potential artefacts are flagged where flux is not associated with an astronomical source if the source is only detected in one optical/NIR band, if the source's Y-band R50 is less than half a pixel, or if the source has an r-Z colour $< -0.75$\,mag (an unphysical colour).
Masking is performed to remove ghosting around bright stars as this is a significant problem in the DEVILS imaging (see figure~8 of \citealt{DaviesDeepExtragalacticVIsible2021}).

This results in 494,084 objects, of which 24,099 have spectroscopic redshifts, 7,307 have grism redshifts and the remaining 462,678 have photometric redshifts. 
\cite{ThorneDeepExtragalacticVIsible2021} demonstrate that \textsc{ProSpect} works reasonably with photometric redshifts, so we use all available redshifts. 

\begin{figure*}
    \centering
    \includegraphics[width = \linewidth]{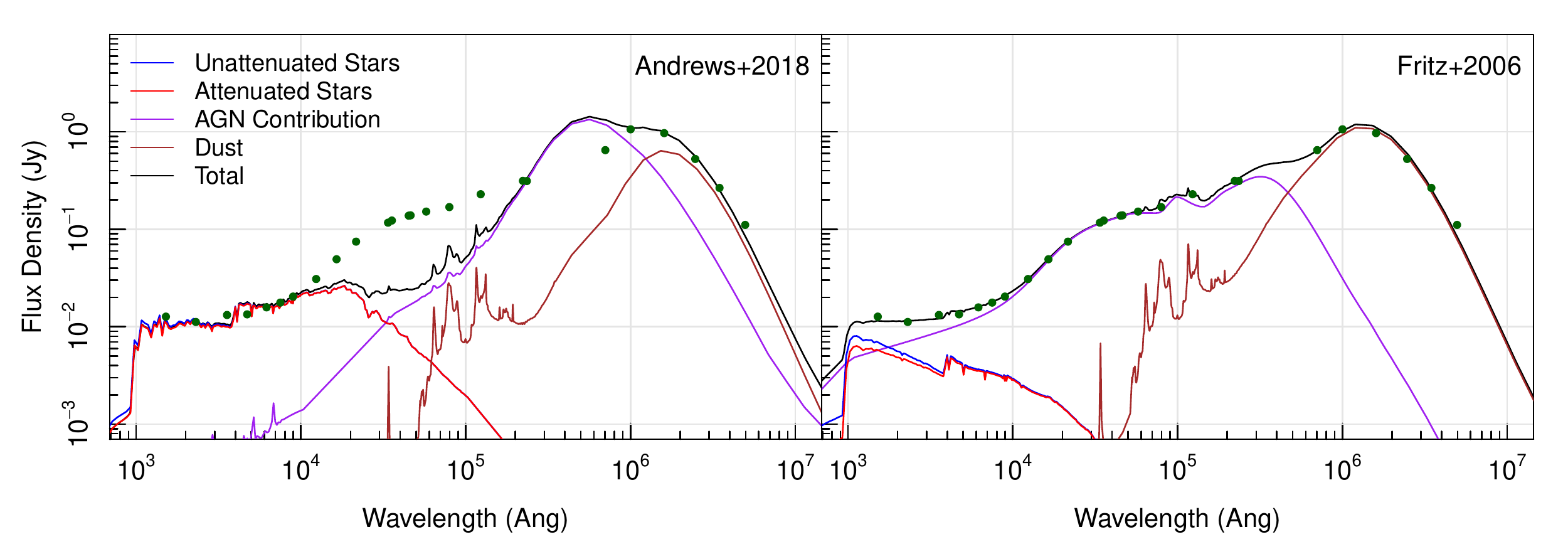}
    \caption{The results of fitting Ark 120 with \textsc{ProSpect} when using the AGN template used in \citet{AndrewsModellingcosmicspectral2018} (left) and the AGN model used in this work and described by \citet{FritzRevisitinginfraredspectra2006,FeltreSmoothclumpydust2012} (right).
    The different SED components are shown including the unattenuated and attenuated stellar emission (blue and red respectively), the AGN contribution (purple), the re-emission from dust (brown) and the total resulting SED (black). 
    We show the input data as the green points with the observed wavelength shown.}
    \label{fig:3C273}
\end{figure*}

\subsection{Galaxy and Mass Assembly Survey}
We also use the spectroscopic and photometric data from the GAMA survey.
GAMA was a large spectroscopic campaign on the Anglo-Australian Telescope targeting five fields (G02, G09, G12, G15, and G23), amounting to a total sky area of 230 square degrees.
GAMA targets were selected by size and colour above a magnitude limit of $r_\text{mag} \le 19.8$ (or $i_\text{mag} \le 19.0$ in G23), and the survey gathered redshifts for $\sim300,000$ galaxies.

We use the new far-UV to far-IR photometry derived using the \textsc{ProFound} source-finding software \citep{RobothamProFoundSourceExtraction2018} and described in detail by \cite{BellstedtGalaxyMassAssembly2020a}.
The photometric bands from this data release include GALEX \textit{FUV} and \textit{NUV}; VST \textit{u, g, r, i}; VISTA \textit{Z, Y, J, H, K$_S$}; WISE \textit{W1, W2, W3, W4;} and Herschel P100, P160, S250, S350, and S500.

The photometry for GAMA is derived in much the same way as for DEVILS, with minor changes due to differences in depth between the surveys. 
Similarly to DEVILS, the GAMA photometry also covers the FUV-FIR regime, but the biggest difference in the photometry between DEVILS and GAMA is in the MIR.
In GAMA the MIR is provided by the  Wide-field Infrared Survey Explorer (\textit{WISE}) whereas in DEVILS the MIR is covered by Spitzer-IRAC and MIPS. 
For the GAMA photometry the \textsc{ProFound} segmentation mode is used for the \textit{FUV-W2} imaging and the \texttt{FitMagPSF} mode is used for \textit{W3-S500}.
Due to differences in wavelength coverage of the two instruments, the DEVILS photometry includes additional bands between $5-8\,\mu$m whereas the GAMA photometry provides constraint at $12\,\mu$m.
This is not expected to impact the recovered AGN luminosities or host galaxy properties. 
As GAMA is a much shallower survey than DEVILS, the far-infrared imaging in the GAMA fields has sufficient resolution and depth that it can be extracted for all optically detected sources.

For this work we use the three equatorial fields (G09, G12, and G15) as well as G23.
We select all objects with $z>0$, a redshift quality flag $nQ \ge 3$, and an \textsc{UBERCLASS}=\texttt{galaxy} based on size and colour.
This results in a sample of 233,762 galaxies all with spectroscopic redshifts.

\section{SED Modelling}\label{sec:SED}

\subsection{Galaxy Component}
We implement the same method as outlined in \cite{ThorneDeepExtragalacticVIsible2021} for the stellar and dust components, using the photometry presented in \cite{DaviesDeepExtragalacticVIsible2021} and passed into the \textsc{ProSpect} SED fitting code \citep{RobothamProSpectgeneratingspectral2020}.
For a detailed description of the fitting we direct the reader to \cite{ThorneDeepExtragalacticVIsible2021}, however we provide a brief summary in this section.
We use the \cite{BC03} stellar templates, assume a \cite{ChabrierGalacticStellarSubstellar2003} initial mass function and model the dust attenuation and re-emission using the \cite{CharlotSimpleModelAbsorption2000} and \cite{DaleTwoParameterModelInfrared2014} models respectively assuming energy balance. 
We also include a 10 per cent error floor across all bands to account for offsets between facilities and instruments. 
In our analysis we use the \texttt{massfunc\_snorm\_trunc} parameterisation for the star formation history which takes the form of a skewed Normal distribution in lookback time, with the peak position (\texttt{mpeak}), peak star formation rate (\texttt{mSFR}), SFH width (\texttt{mperiod}), and skewness (\texttt{mskew}) set as free parameters. 
The SFH is anchored to 0 at a lookback time of 13.4 Gyr, selected to be the age at which galaxies start forming.

In \cite{ThorneDeepExtragalacticVIsible2021} we emphasise the importance of implementing a physically motivated evolving gas phase metallicity as it results in a systematic 0.2\,dex offset in the recovered stellar masses.
This is also investigated and justified in \cite{RobothamProSpectgeneratingspectral2020} with comparison to the Shark semi-analytic model \citep{LagosSharkintroducingopen2018}.
To do this within \textsc{ProSpect} we map the metallicity evolution of each galaxy linearly to the stellar mass evolution, given by the \texttt{Zfunc\_massmap\_lin} function. 
This ensures that chemical enrichment in the galaxies follows the assumed star formation rate (SFR), where increased star formation is associated with an increased rate of metal production.
The final metallicity for each galaxy is allowed to be a free parameter, \texttt{Zfinal}.
We highlight that this value represents the gas-phase metallicity at observation of the object, and correspondingly the metallicity of the youngest stars in the galaxy. 
Using the recovered star formation and metallicity histories of a sample of 7,000 galaxies with $z<0.06$, 
\cite{BellstedtGalaxyMassAssembly2020b,BellstedtGalaxyMassAssembly2021} show that this simple yet physical assumption allows for an accurate recovery of the cosmic star formation history (CSFH) and evolution of the mass-metallicity relation. 
The specific impact of various assumed metallicity implementations on the CSFH was demonstrated in figure 4 of \cite{BellstedtGalaxyMassAssembly2020b}.

In addition to the five free parameters specifying the star formation and metallicity histories, we include four free parameters to describe the contribution of dust to the SED. 
Within \textsc{ProSpect} the dust is assumed to exist in two forms; in birth clouds formed around young stars, or distributed as a screen in the interstellar medium. 
For each of these components we include two free parameters, describing the dust opacity (\texttt{tau\_screen}, \texttt{tau\_birth}), and the dust radiation field intensity (\texttt{alpha\_screen}, \texttt{alpha\_birth}). 
See figure 3 of \cite{ThorneDeepExtragalacticVIsible2021} for the impact of each parameter on a generated galaxy SED.

\subsection{AGN Component}\label{sec:AGNComp}

\begin{figure}
    \centering
    \includegraphics[width = \linewidth]{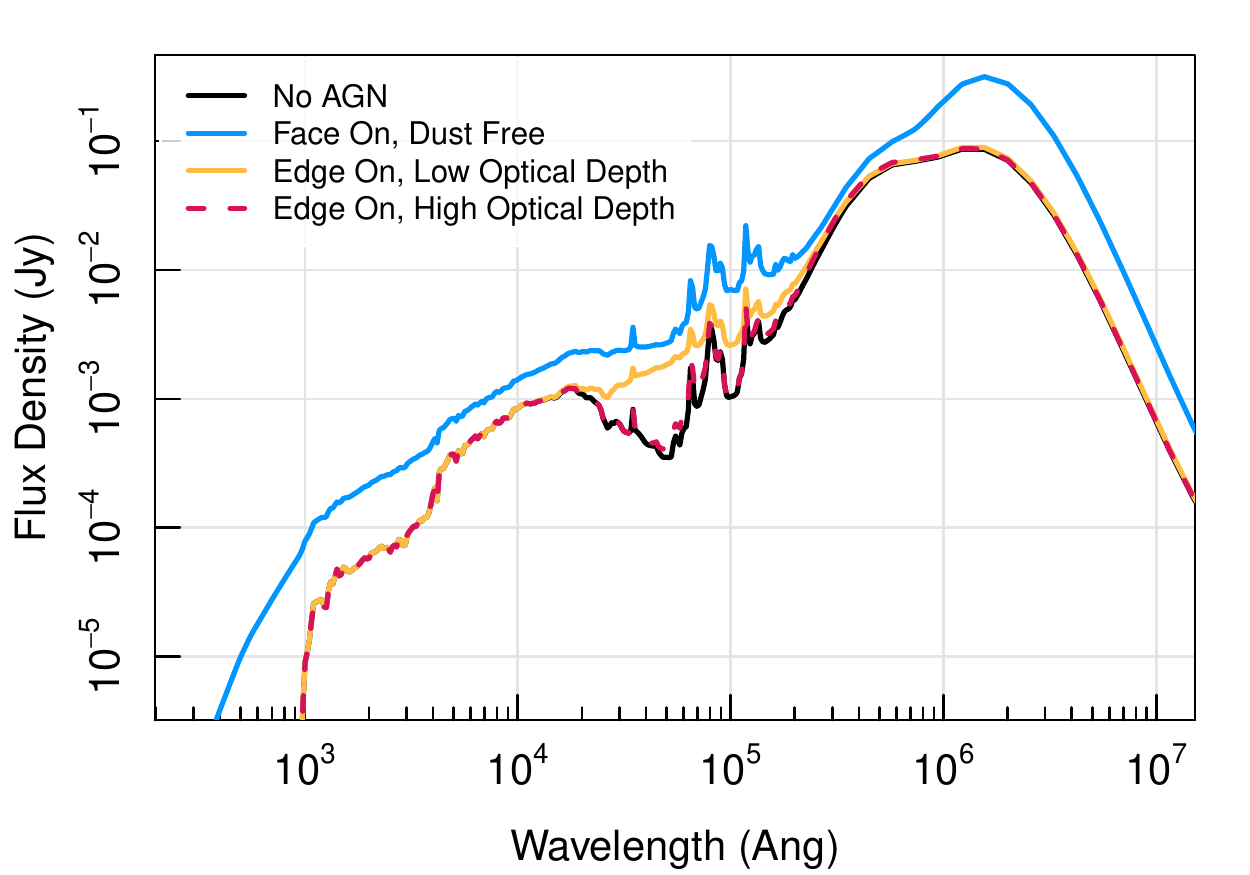}
    \caption{The resulting SED for a simulated galaxy with constant star formation and total stellar mass formed of $10^{10} M_\odot$.  We show the SED with no AGN component in black, and an AGN component with central engine power of $\texttt{AGNlum} = 10^{44}\,\text{erg s}^{-1} $  shown in a face-on orientation (\texttt{AGNan} = $90^{\circ}$) is blue, and in an edge on orientation (\texttt{AGNan} = $0^{\circ}$) with a low torus dust optical depth (\texttt{AGNta} = 1) in yellow, and a high optical depth (\texttt{AGNta} = 10) in dashed red.
}
    \label{fig:AGNSEDExamples}
\end{figure}

\begin{figure*}
    \centering
    \includegraphics[width = \linewidth]{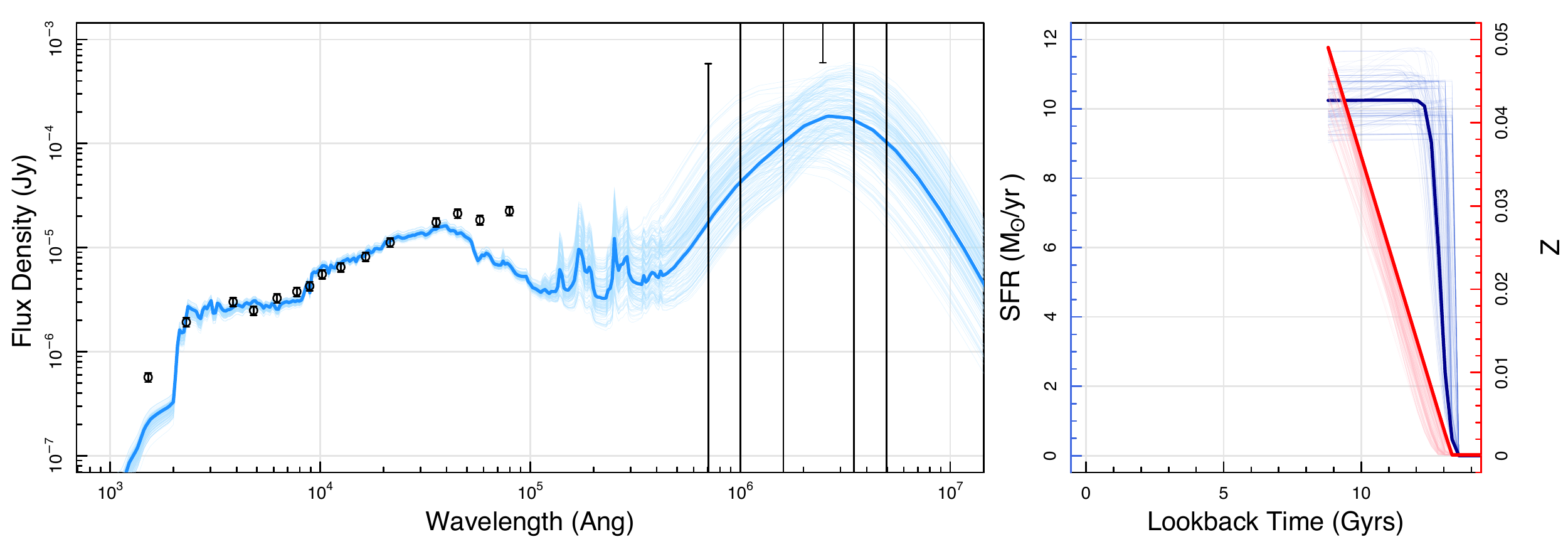}
    \includegraphics[width = \linewidth]{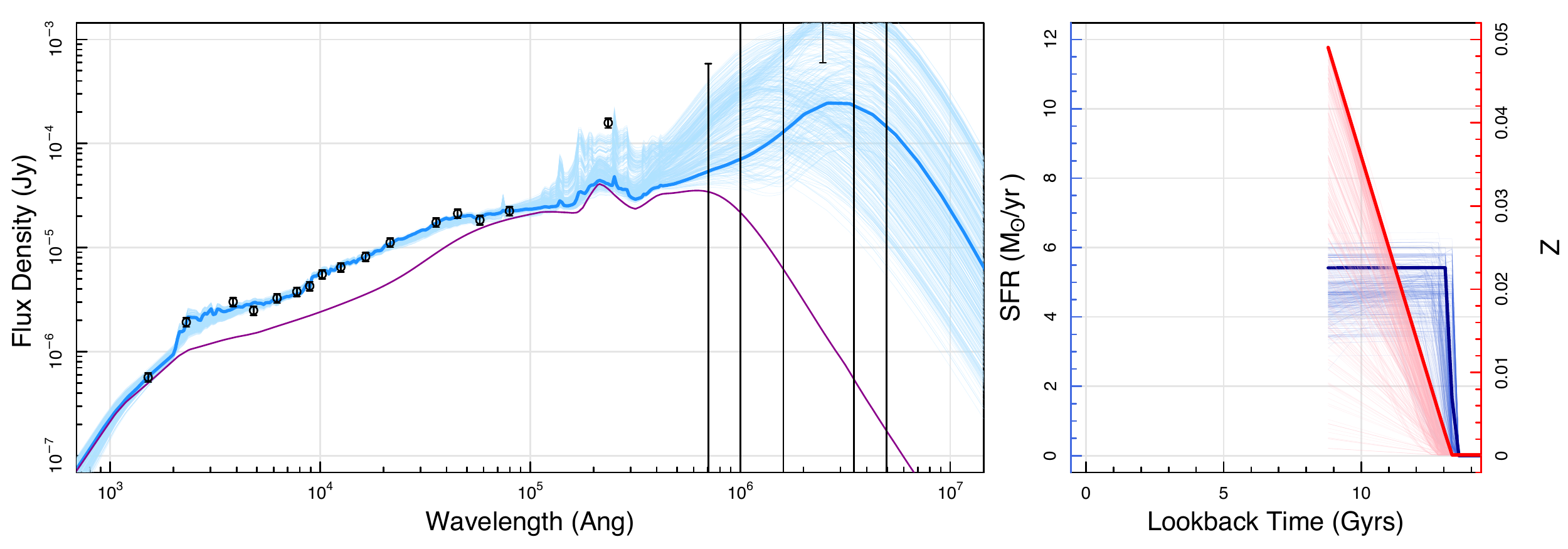}
    \caption{Here we present example SED outputs as a function of observed wavelength for galaxy 101494591582430080 at $z = 1.24$ fit without and with an AGN component in the upper and lower panels respectively. 
    The MIPS24 data point is not shown in the upper panel as bands within the PAH features were removed from the fitting when an AGN component was not included (see \citealt{ThorneDeepExtragalacticVIsible2021}). 
    We show the contribution from the AGN component in the lower panel as the purple line. 
    The left panels show the input photometry (black circles and error bars), the best fit SED in blue and the SED generated from each step of the final MCMC chain in pale blue. 
    The right panels show the best fit SFH in blue and the posterior sampling in pale blue with the scaling given by the left axis.
    The metallicity history for each galaxy is also presented in the right panel as the red line for the best fit solution and as the pink lines for sampling of the posterior. 
    The scale of the metallicity history is shown on the right axis. 
   }
    \label{fig:SED_Example}
\end{figure*}

\begin{figure}
    \centering
    \includegraphics[width = \linewidth]{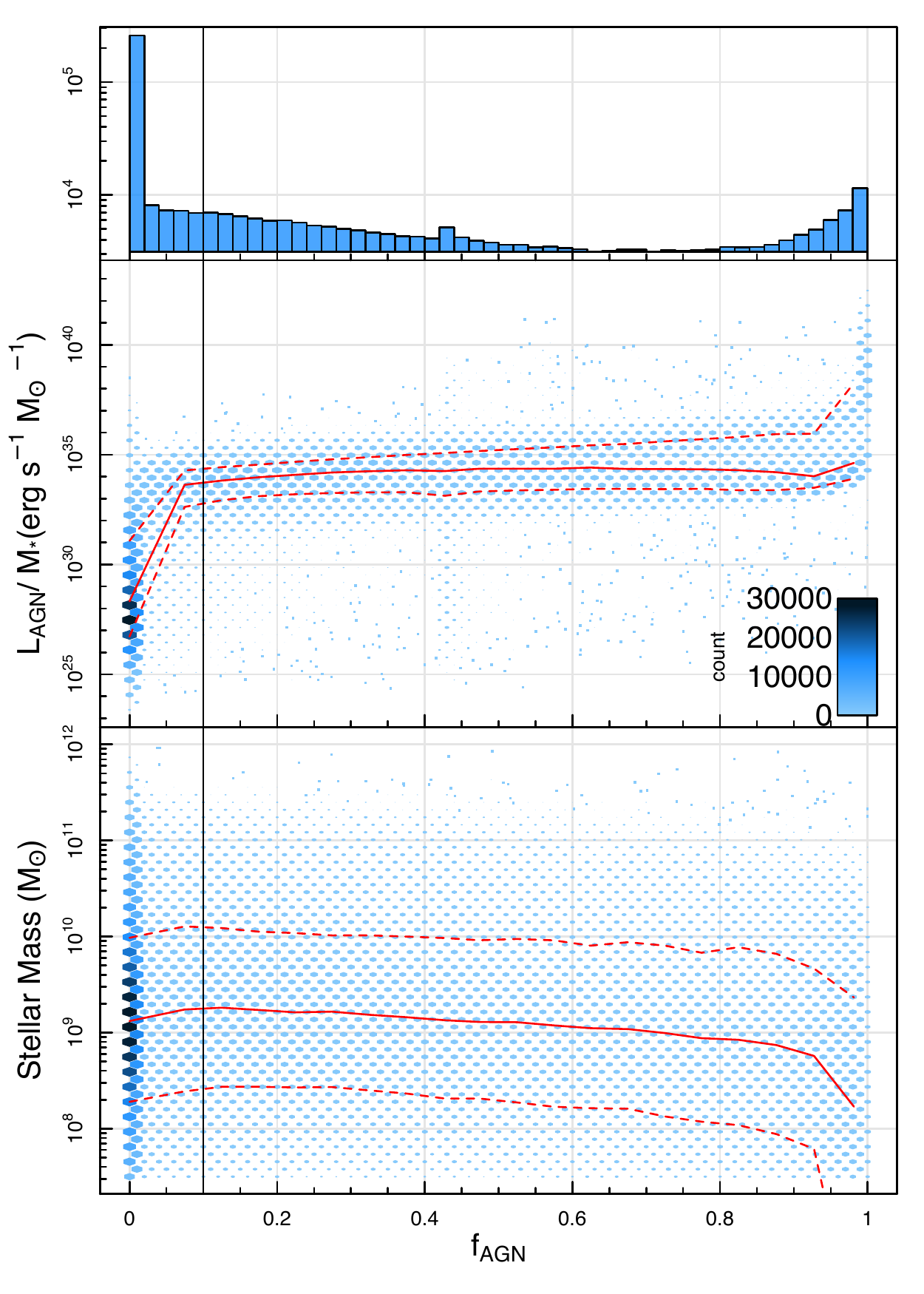}
    \caption{
    The distribution of \textsc{ProSpect}-derived $f_\text{AGN}$ values for the full DEVILS D10 sample \textit{(top)}. 
    Note the increasing frequency of objects with $f_\text{AGN} > 0.8$ is primarily due to cases where the AGN component entirely dominates the SED with minimal contribution from the host galaxy due to lack of FIR photometry.
    Density distribution of the specific AGN luminosity ($L_\text{AGN}$/$M_\star$) \textit{(middle)} and stellar mass \textit{(bottom)} as a function of $f_\text{AGN}$ for all sources in the DEVILS-D10 sample. 
    Both the size and colour reflect the number of sources in each bin.
    We show the running median and one sigma ranges as the solid and dashed red lines respectively. 
    The vertical black line shows the $f_\text{AGN} > 0.1$ threshold used in this work.
    }
    \label{fig:fAGN_sAGN}
\end{figure}

\begin{table*}
    \centering
    \caption{The parameters used by \textsc{ProSpect} in this work. We list the parameter name, a brief description, whether it is fit in linear or logarithmic (log) space or if it is fixed, the range of allowed values and any imposed non-uniform prior.}
    \label{tab:Parameters}
    \begin{tabular}{l l l l l l}
    \hline
    Parameter & Description & Type & Units & Values & Prior \\
    \hline
    \texttt{mSFR} &  peak star formation rate &  log & $M_\odot\,\text{yr}^{-1}$ & [-3,4] & \\
    \texttt{mpeak} &  lookback time when peak star formation occurred & linear & Gyr & [-2,13.38] & \\
    \texttt{mperiod} &  width of the SFH & log &  Gyr & [$\log_{10}(0.3)$,2] & $100 \erf (\texttt{mperiod}+2) - 100$ \\
    \texttt{mskew} &  skewness of the SFH & linear  & & [-0.5,1] & \\
    \texttt{Zfinal} &  final gas-phase metallicity & log & & [-4, -1.3] &\\  
    \hline
     \texttt{alpha\_SF\_birth} &  Power law of the radiation field heating birth cloud dust  & linear & & [0,4] & $\exp{(-\frac{1}{2} (\frac{\alpha_\text{birth} + 2}{1})^2 ) }$ \\
    \texttt{alpha\_SF\_screen} &  Power law of the radiation field heating general ISM dust  & linear & & [0,4] & $\exp{(-\frac{1}{2} (\frac{\alpha_\text{screen} + 2}{1})^2 ) }$ \\
       \texttt{tau\_birth} & optical depth of the birth clouds & log & & [-2.5,1] & $\exp{(-\frac{1}{2} (\frac{\tau_\text{birth} - 0.2}{0.5})^2 ) }$ \\
    \texttt{tau\_screen} &  optical depth of the general ISM & log & & [-5,1]  & $-20\erf(\tau_\text{screen}-2) $  \\
    \hline
     \texttt{AGNan} & angle of observation & linear & deg & [0.001,89.990] & \\
         \texttt{AGNlum} & bolometric luminosity of AGN source & log & erg s$^{-1}$ & [35,49] & \\
           \texttt{AGNta} & optical depth tau & log & & [-1,1] & \\
        \texttt{AGNrm} &  outer to inner torus radius ratio & fixed & & 60 &\\
         \texttt{AGNbe} & beta dust parameter & fixed & & -0.5 &  \\
         \texttt{AGNal} & gamma dust parameter & fixed  & & 4.0 & \\
         \texttt{AGNct} & opening angle of torus & fixed & deg & 100 & \\
    \hline
    \end{tabular}
\end{table*}

Three AGN templates/models have, at this stage, been made available within \textsc{ProSpect}; the one contained within \cite{DaleTwoParameterModelInfrared2014}, the extended AGN template from \cite{AndrewsModellingcosmicspectral2018} and the model initially presented in \cite{FritzRevisitinginfraredspectra2006} and expanded in \cite{FeltreSmoothclumpydust2012}.
Figure~\ref{fig:3C273} shows the SED of Ark 120 with photometry from \cite{Brownspectralenergydistributions2019}, fit with both the template used in \cite{AndrewsModellingcosmicspectral2018} and the more flexible model described in \cite{FritzRevisitinginfraredspectra2006} and \cite{FeltreSmoothclumpydust2012}. 
It is clear from Figure~\ref{fig:3C273} that the \cite{AndrewsModellingcosmicspectral2018} template does not have the required flexibility to fit the MIR excess while the \cite{FritzRevisitinginfraredspectra2006} model does. 
While this example shows the benefits of the flexibility of the \cite{FritzRevisitinginfraredspectra2006} template in the MIR, in other cases we have found that the larger wavelength coverage into the ultraviolet regime by the \cite{FritzRevisitinginfraredspectra2006} models also allows for greater constraint of the AGN parameters.

\cite{FritzRevisitinginfraredspectra2006} models the primary AGN source as a composition of power-laws, with different spectral indices as a function of the wavelength. 
The contribution of the power-law in the UV-optical can have a considerable impact on the resulting galaxy properties \citep{CardosoImpactAGNfeatureless2017}.
As such, it is important to model the contribution from the primary source in addition to the emission from the torus in the MIR. 
To model the contribution from the torus, \cite{FritzRevisitinginfraredspectra2006} uses a simple but realistic torus geometry, a flared disc, and a dust grain distribution function including a full range of grain sizes and assumes that the dust in the AGN torus is smoothly distributed. 
There has been much debate as to whether the dust in an AGN torus is smoothly distributed \citep{PierInfraredspectraobscuring1992, DullemondClumpytoriactive2005, FritzRevisitinginfraredspectra2006,FranceschiniCosmicevolutiongalaxy2006}, clumpy \citep{KrolikMoleculartoriSeyfert1988, NenkovaDustEmissionActive2002,NenkovaAGNDustyTori2008a, ElitzurAGNobscuringTorusEnd2006,TristramResolvingcomplexstructure2007}, or a combination of the two \citep{Stalevski3Dradiativetransfer2012, AssefMidInfraredSelectionActive2013}.
While observations of the strength of the silicate feature at $9.7\,\mu$m in AGN seem to favour models where the dust is predominantly clumpy, \cite{FeltreSmoothclumpydust2012} argued that observations are not yet able to discriminate between the different models. 
\cite{FeltreSmoothclumpydust2012} also conclude that the properties of dust in AGN as measured by model fitting will strongly depend on the choice of the dust distribution.
As such, we continue to treat dust as a nuisance parameter but include it when fitting to allow for more accurate estimates of the AGN luminosity.
The torus dust is modelled as a combination of graphite and silicate particles with a power-law distribution of sizes with scattering and absorption coefficients taken from \cite{LaorSpectroscopicconstraintsproperties1993}. 
The model from \cite{FritzRevisitinginfraredspectra2006} facilitates a more detailed fitting of the torus geometry and physics including the outer to inner torus radius, the opening angle of the torus, and the angle of observation (where $\theta = 0^{\circ}$ is edge-on through the torus, and $\theta = 90^{\circ}$ is polar aligned). 
Within \textsc{ProSpect} the emission from the central source and dust torus is also re-attenuated through the general ISM screen (see figure 1 of \citealt{RobothamProSpectgeneratingspectral2020}).

Within \textsc{ProSpect} we model the AGN contribution by allowing the bolometric luminosity of the central source (\texttt{AGNlum}), optical depth at 9.7$\mu$m (\texttt{AGNta}), and angle of observation (\texttt{AGNan}) to vary freely between the limits given in Table~\ref{tab:Parameters}. 
Motivated by \cite{PouliasisobscuredAGNpopulation2020}, we fix the outer to inner torus radius ratio (\texttt{AGNrm}), the parameters controlling the radial and angular distribution of torus dust (\texttt{AGNbe} and \texttt{AGNal}), and the opening angle of the torus (\texttt{AGNct}) to the values given in Table~\ref{tab:Parameters}. 
These values are fixed as there are significant degeneracies between the AGN parameters that are difficult to disentangle even when fitting emission purely from an AGN with no host galaxy contribution. 
Currently \textsc{ProSpect} does not include UV absorption but this is planned for future versions and has no impact on derived AGN luminosities.

Figure~\ref{fig:AGNSEDExamples} shows the impact of various AGN implementations on the overall galaxy SED. 
We show the resulting SED for a galaxy with constant star formation and a total stellar mass formed of $10^{10}\,M_\odot$ as the black line.
We also show the SED of a galaxy with the same stellar mass and star formation history but with an AGN component with a bolometric luminosity $\texttt{AGNlum} = 10^{44}\,\text{erg s}^{-1}$ in a face-on orientation (\texttt{AGNan} = $90^{\circ}$), and two edge-on cases (\texttt{AGNan} = $0^{\circ}$) with varying dust torus optical depths.
The face-on AGN SED differs the most from the SED without an AGN component as the AGN impacts the emission of the galaxy across the entire modelled wavelength range (FUV-FIR). 
The edge on, low optical depth (\texttt{AGNta} = 1) case differs from the no AGN case across the NIR-MIR and can often be identified through rising flux measurements in increasing wavelength IRAC bands. 
The edge on, high optical depth (\texttt{AGNta} = 10) case differs from the no AGN case only in the MIR where there is a slight increase in the flux. 
Note that this wavelength range of the SED is poorly sampled at most redshifts in this sample and can be very hard to constrain. 

In \cite{ThorneDeepExtragalacticVIsible2021}, we removed wavelength bands that fall within the polycyclic aromatic hydrocarbon (PAH) features (restframe wavelengths of $5-15\,\mu$m) when fitting due to large residuals at these wavelengths when compared to the \cite{DaleTwoParameterModelInfrared2014} dust templates. 
For some AGN cases, these bands provide the constraint of an AGN component and cannot be ignored for this work.
We therefore include the photometric measurements in these wavelength bands in this AGN-focused work.
We also use a longer optimisation routine for fitting AGN than that used in \cite{ThorneDeepExtragalacticVIsible2021} as the degeneracies introduced by the AGN model increase the prevalence of local minima in the fitted parameter space, requiring increased sampling to identify the true global minimum.
For this work we use the \textsc{Highlander} R package\footnote{\url{https://github.com/asgr/Highlander}}, which was originally developed for use with \textsc{ProFit} \citep{profit}, and alternate between genetic optimisation and Markov Chain Monte Carlo (MCMC) phases twice with 2000 iterations in each case for 8000 iterations total. 
This is 6000 steps longer than in \cite{ThorneDeepExtragalacticVIsible2021} and takes approximately 15 minutes on a modern CPU.
To ensure reasonable fits for all objects, we re-fit objects with a significantly worse fit with an AGN component than without ($\Delta \text{log-likelihood} = 20$), and objects with a significantly bad likelihood in general.
Fits with low likelihoods arise due to the optimisation routine not finding the true global minimum, but instead finding local minima which produce solutions with large residuals.
These objects are re-fit using an extra 1000 steps in each of the genetic optimisation phases. 
After re-fitting these objects, all 494,084 galaxies have secure and accurate SED fits.
We present each of the parameters used by \textsc{ProSpect}, a brief description and their values in Table~\ref{tab:Parameters}.

\subsection{Outputs}
We show the SED of an example galaxy from the DEVILS sample with significant AGN contribution to the MIR emission (Figure~\ref{fig:SED_Example}) fit with no AGN component (as per \citealt{ThorneDeepExtragalacticVIsible2021}) in the top panel and with an AGN component in the bottom panel. 
When fitting with an AGN component, the derived stellar mass and SFR of this galaxy are lower by 0.13 and 0.27\,dex respectively than when no AGN component is included (see Section~\ref{sec:EffectOnHostGalaxies}). 
The fit in the MIR and FUV is significantly better for this galaxy when including an AGN component which is reflected in the likelihood of the fit.
Note that these are rare cases in the results from \cite{ThorneDeepExtragalacticVIsible2021} where most fits have small residuals in the FUV and MIR. 

As an indication of the performance of our technique we include the \textsc{ProSpect} fits to the SEDs of well known low-redshift AGN from \cite{Brownspectralenergydistributions2019} in Appendix~\ref{app:Brown}.
The remainder of this work explores the ability to identify and quantify AGN in the DEVILS D10 field using \textsc{ProSpect} as described above.

In order to classify galaxies as containing a significant AGN component we define the value $f_\text{AGN}$, which is calculated to be the fraction of flux contributed by the AGN component between $5-20\,\mu$m (based on the definition in \citealt{DaleTwoParameterModelInfrared2014}). 
We define a significant AGN contribution as objects that have $f_\text{AGN} > 0.1$ as used by \cite{LejaHotDustPanchromatic2018}.
The value of $f_\text{AGN}$ is calculated after fitting from the best fit total and AGN component SEDs.

Figure~\ref{fig:fAGN_sAGN} shows the distribution of recovered $f_\text{AGN}$ values, how they vary with galaxy stellar mass ($M_\star$), and with what we are defining as specific AGN luminosity ($L_\text{AGN}$/$M_\star$). 
The increased number of objects with $f_\text{AGN} > 0.8$ is primarily due to cases where the AGN component entirely dominates the SED with minimal contribution from the host galaxy due to lack of FIR photometry.
As the ability for \textsc{ProSpect} to identify AGN depends on both the luminosity of the host galaxy and the luminosity of the AGN source, we expect objects with low specific AGN luminosity to also have low $f_\text{AGN}$.
Objects with a significant AGN contribution can be seen as the horizontal band across the figure, whilst objects with no significant AGN contribution can be seen on the left of the figure spanning a larger range of specific AGN luminosity values. 
The rising median $L_\text{AGN}$/$M_\star$ value for $f_\text{AGN} < 0.1$ supports the use of the $f_\text{AGN} > 0.1$ threshold to isolate galaxies with significant AGN contribution. 

We show there is no clear trend in stellar mass as a function of $f_\text{AGN}$ which demonstrates that our recovered $f_\text{AGN}$ are not biased towards certain stellar mass ranges. 
For $f_\text{AGN} > 0.9$, there is a slight turn down in median stellar mass. 
As described above, this is due to cases where the AGN component dominates the SED due to poor constraint in the FIR and the resulting stellar mass is unconstrained. 
These objects are generally at very high redshift ($z>4$) and are not used for comparisons in Section~\ref{sec:comparisons} or for the calculation of the luminosity function in Section~\ref{sec:AGNLF}.

\textsc{ProSpect} recovers $f_\text{AGN} > 0.1$ for 205,668 objects in the DEVILS sample (41.5 per cent) and 67,258 galaxies in the GAMA sample (28.7 per cent). 
The higher fraction of AGN in the DEVILS sample is because of two reasons. 
Firstly, as DEVILS covers a higher redshift range than GAMA, we expect the fraction of objects that host an AGN to increase with redshift. 
Secondly, as the FIR imaging in the DEVILS field is not deep enough to extract photometry for all objects, especially objects that are faint or at high redshift, there is minimal constraint on an AGN component. 
This means that in some instances, a significant AGN component ($f_\text{AGN} > 0.1$), can be included in the MIR-FIR without impacting the host galaxy contribution to the FUV-NIR or the likelihood of the fit. 
In GAMA, FIR photometry is extracted for every object with a spectroscopic redshift resulting in FIR constraint for every source used in this work.

\begin{figure}
    \centering
    \includegraphics[width = \linewidth]{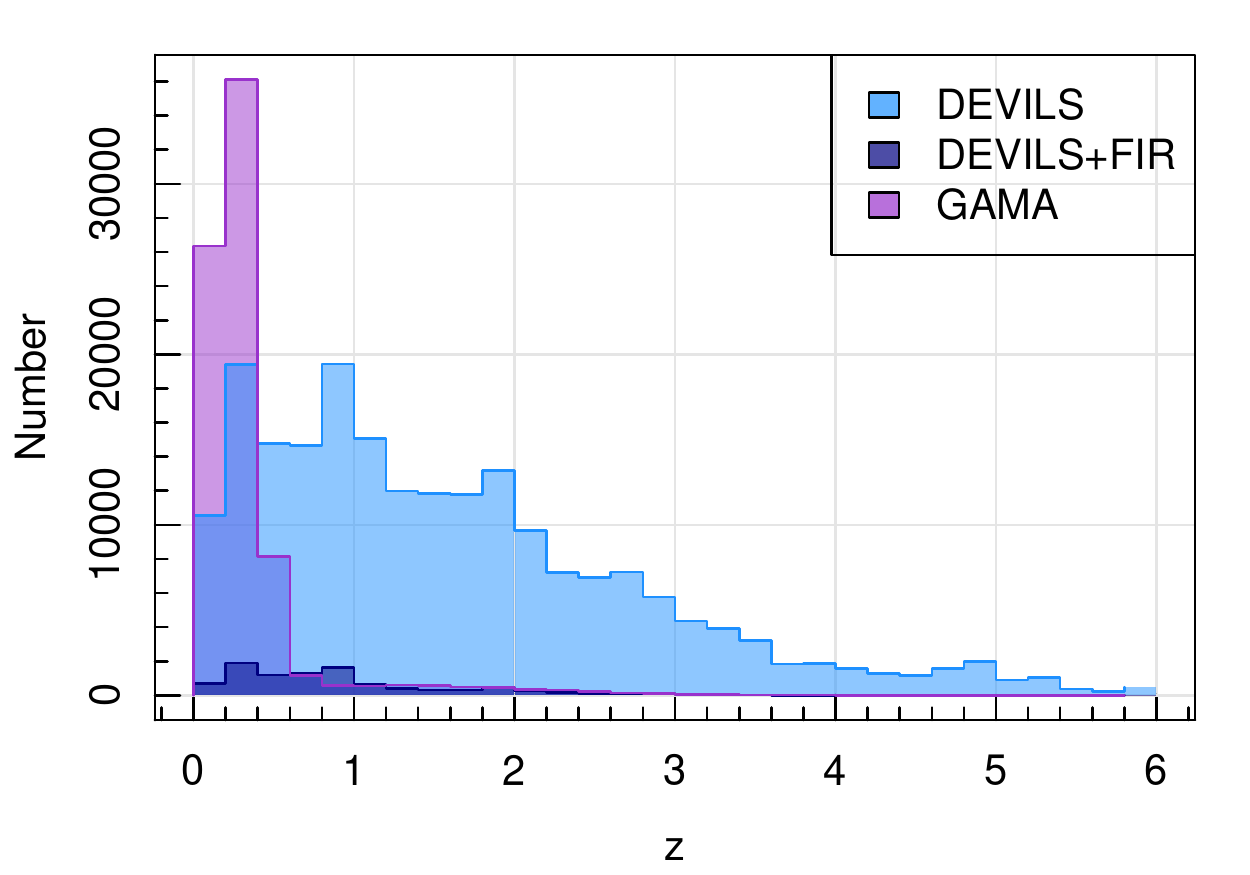}
    \caption{The redshift distribution of the recovered AGN ($f_\text{AGN} > 0.1$) from DEVILS (light blue) and GAMA (purple). We also show the sub-sample of AGN from DEVILS that have measured FIR photometry (dark blue).}
    \label{fig:RedshiftDist}
\end{figure}

Excluding objects with no FIR constraint we recover 9,761 AGN in the DEVILS sample.
These objects span $0 < z < 6$, but 90 per cent of these objects have $z < 2$. 
We show the redshift distribution of the AGN samples in Figure~\ref{fig:RedshiftDist}.
For the remainder of this work we use the full sample of AGN recovered in DEVILS as although the AGN component may not be well constrained, the inclusion of a significant AGN component can impact the derived host galaxy properties.

\section{Comparisons to other AGN identification methods}\label{sec:comparisons}

\begin{table*}
    \centering
    \caption{
    Summary of the AGN identification comparisons. 
    We list the method used, any selections applied to the total galaxy sample and the particular criteria used. 
    In each case we also include the number of objects that meet the literature criteria, the number of objects that have a \textsc{ProSpect}-derived $f_\text{AGN} > 0.1$, and the number of objects that satisfy both the literature criteria and have $f_\text{AGN} > 0.1$.
    We also include the percent of objects that satisfy both selections given that they satisfy the criteria (\% Criteria) or are selected as AGN by \textsc{ProSpect} (\% \textsc{ProSpect}).}
    
    \label{tab:MethodSummary}
    \begin{tabular}{l l l c c c c c }
    \hline
Method	&	Selection	&	Criteria	&	N Criteria 	&	N \textsc{ProSpect} &	N Both 	 &	\% \textsc{ProSpect}	&	\% Criteria	 \\
	&		&		&	 $N_C$	&		$N_P$&	 	$N_B$ &	$N_B / N_C$   &		$N_B / N_P$ \\
\hline									
X-ray	&		&	2 arcsec position match to	&	2,345	&	9,603	&	1,528	&	65.2	&	15.9	\\
(DEVILS) & &  \citet{MarchesiCHANDRACOSMOSLEGACY2016} catalogue & & & & & \\
\hline
BPT	&	S/N > 3 in each emission line	&	\citet{KewleyTheoreticalModelingStarburst2001}	&	2,404	&	7,062	&	936	&	38.9	&	13.3	\\
(GAMA)	&	total N = 34,571	&	\citet{Kauffmannhostgalaxiesactive2003}	&	7,843	&	7,062	&	1,671	&	21.3	&	23.7	\\
	&		&	Seyfert + Broad Lines	&	163	&		&	149	&	91.4	&		\\
\hline	
Spitzer 	&	S/N > 5 in all IRAC bands	&	\citet{LacySpitzerMidinfraredActive2013}	&	45,929	&	57,804	&	24,944	&	54.3	&	43.2	\\
(DEVILS)	&	total N = 116,449	&	\citet{DonleyIdentifyingLuminousActive2012}	&	14,944	&	57,804	&	10,309	&	69.0	&	17.8	\\
\hline															
WISE	&	S/N > 5 in W1 and W2	&	\citet{SternMIDINFRAREDSELECTIONACTIVE2012}	&	4,137	&	67,036	&	3,386	&	81.8	&	5.1	\\
(GAMA)	&	total N = 232,076	&	\citet{AssefWISEAGNCatalog2018} R90	&	8,385	&	67,036	&	5,932	&	70.7	&	8.8	\\
	&		&	\citet{AssefWISEAGNCatalog2018} R75	&	13,364	&	67,036	&	8,167	&	61.1	&	12.2	\\
	&		&	\citet{AssefWISEAGNCatalog2018} C90	&	38,334	&	67,036	&	16,352	&	42.7	&	24.4	\\
	&		&	\citet{AssefWISEAGNCatalog2018} C75	&	17,020	&	67,036	&	9,407	&	55.3	&	14.0	\\
\hline
    \end{tabular}
\end{table*}

AGN have unique observational signatures over the entire electromagnetic spectrum. 
Although many types of AGN exist in the literature it is becoming increasingly clear that these classifications are only partially related to intrinsic differences between AGN, and primarily reflect variations in the method by which each class of AGN is selected.
Multiple studies show that AGN indicators are far from complete and can pick out different and often non-overlapping AGN populations \citep{JuneauWidespreadHiddenActive2013,TrumpBiasesOpticalLineRatio2015}. 
As the processes driving radio emission from AGN can be uncoupled from the processes that drive emission over the rest of the electromagnetic spectrum, we focus here on X-ray, MIR, and optical selection, deferring detailed radio comparisons to Thorne et al. (in prep.).
In the following sections we compare our SED AGN selection with AGN selected using different approaches to demonstrate the effectiveness of our method.

\subsection{X-ray}\label{sec:xray}
The accretion disc surrounding a black hole is believed to produce a thermal spectrum, with most photons produced at UV-optical wavelengths.
Some of these photons are scattered to higher energies by relativistic photons via inverse Compton processes \citep{HaardtXrayspectratwophase1993} and appear as an approximate power-law shape at X-ray wavelengths.
X-ray identification of AGN is thought to be the least biased as X-rays penetrate through dust and gas efficiently and X-ray emission from host galaxy processes is typically weak when compared to the AGN \citep{BrandtCosmicXraysurveys2015,PadovaniActivegalacticnuclei2017}. 
However, it is possible that $\sim\,40$ per cent of Compton-thick AGN at redshifts $z > 0.5-1.5$ would remain undetected in even the deepest \textit{Chandra} surveys \citep{BrandtXRaySurveyResults2006}.
\cite{JuneauWidespreadHiddenActive2013} find that $\sim 35$ per cent of optical- and IR-selected AGN are undetected in X-ray observations, while other studies find anywhere between $22$ to $50$ per cent of MIR-selected AGN are detected in X-rays \citep{DonleyIdentifyingLuminousActive2012,CowleyZFOURGEcatalogueAGN2016,KossNewPopulationComptonthick2016,IchikawaCompleteInfraredView2017}.

To compare our \textsc{ProSpect}-selected AGN to those selected from X-ray measurements we use data from the \textit{Chandra} COSMOS-Legacy project, a 4.6Ms \textit{Chandra} program covering 2.2 deg$^2$ of the COSMOS field (see \citealt{MarchesiCHANDRACOSMOSLEGACY2016} for more details). 
\textit{Chandra} COSMOS-Legacy is deep enough to identify obscured sources with no clear AGN signatures in the optical spectra or MIR up to $z\approx 6$ and X-ray luminosity $L_X \approx 10^{45}$\,erg s$^{-1}$.
The benefit of using \textit{Chandra} for AGN identification is that it has sub-arcsecond resolution imaging providing the best possible source positions with no significant source confusion \citep{BrandtXRaySurveyResults2006}.

Using the catalogue described in \cite{MarchesiCHANDRACOSMOSLEGACY2016}, we perform a position match to our catalogue within a 2 arcsecond radius using the \texttt{coordmatch} function in the \textsc{Celestial} R package \citep{RobothamCelestialCommonastronomical2016}.
We find X-ray counterparts for 2345 of our sources but note that 35 per cent of these objects are not selected as AGN by \textsc{ProSpect}. 
This is not surprising as \cite{MerloniEvolutionactivegalactic2013} find that at $L_\text{bol} = 10^{45} \text{erg s}^{-1}$, hard X-ray samples miss up to 30 per cent of objects while MIR colour selections miss $\approx 60-80$\, per cent. 

\begin{figure}
    \centering
    \includegraphics[width = \linewidth]{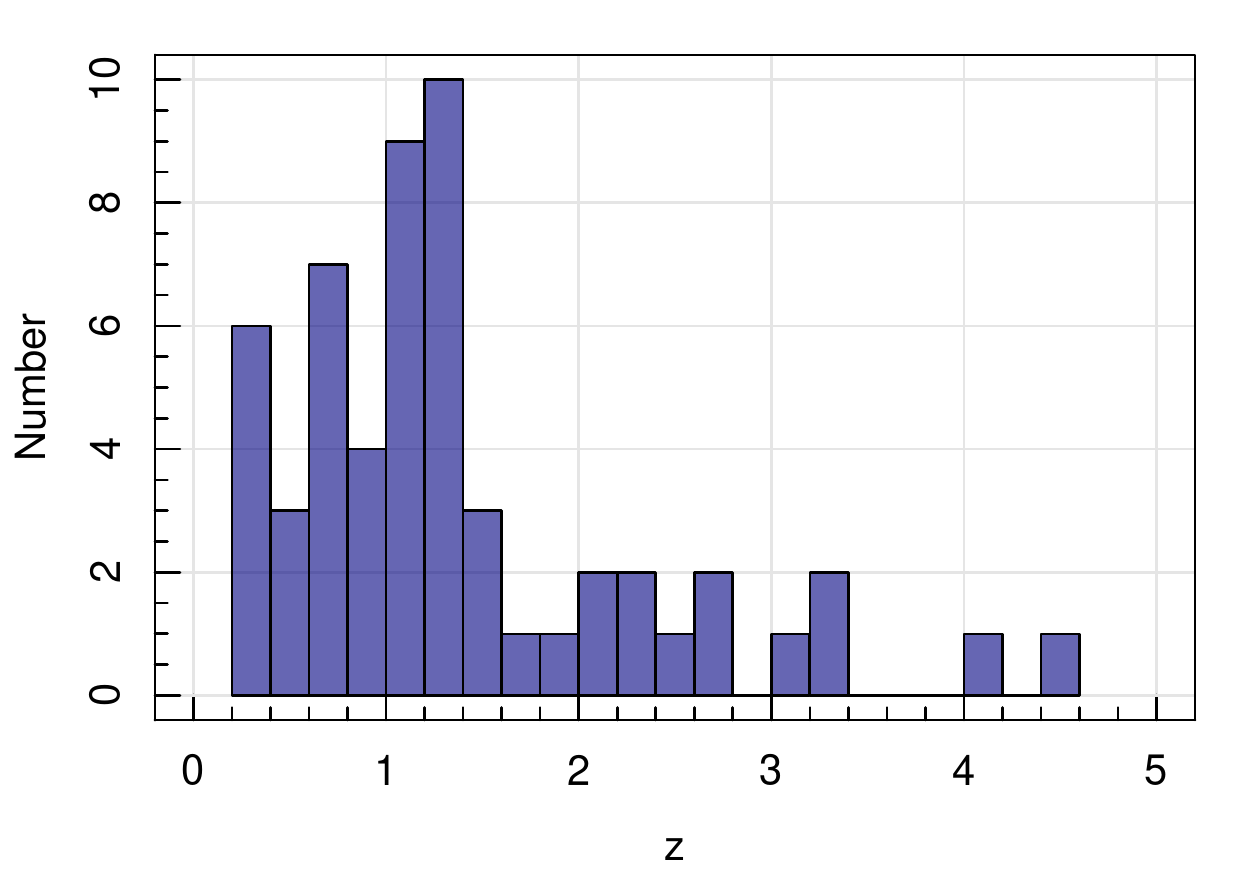}
    \caption{The redshift distribution of the zCOSMOS broad line AGN fit with \textsc{ProSpect}.}
    \label{fig:BLAGN}
\end{figure}

Objects with X-ray luminosities $L_\text{2-10keV}^\text{X-ray} < 10^{42}$\,erg s$^{-1}$ could be contaminated by emission from X-ray binaries and/or hot interstellar medium gas. 
There are 32 objects in this sample that have $L_\text{bol}^\text{X-ray} < 10^{43}$\,erg s$^{-1}$, so to ensure the sample is not contaminated, we use the X-ray hardness ratio for each object calculated using Bayesian Estimation of Hardness Ratios (BEHF; \citealt{ParkBayesianEstimationHardness2006}) and presented in \citet{CivanoChandraCosmosLegacy2016} and \citet{MarchesiCHANDRACOSMOSLEGACY2016}.
The hardness ratio is defined as follows;
\begin{equation}
    \text{HR} = (H - S)/(H + S),
\end{equation}

where $H$ and $S$ are the count rates in the 0.5–2 keV and 2–10 keV X-ray bands.
Using the hardness ratio as a proxy for X-ray emission type, we confirm that all sources have a hardness ratio above that of thermal emission from X-ray binaries or hot interstellar medium gas (i.e. $\text{HR} > - 0.8$), supporting classification as AGN \citep{MezcuaIntermediatemassblackholes2018}.

\textsc{ProSpect} identifies an AGN in 65 per cent of X-ray detected sources (1,528 objects), which we confirm have X-ray emission consistent with that of an AGN.
We do find that \textsc{ProSpect} predicts an X-ray counterpart detectable above the \textit{Chandra} sensitivity limit for $\sim8,000$ additional objects with $f_\text{AGN} > 0.1$.
These AGN could be Compton-thick, and could remain undetected in even the deepest \textit{Chandra} surveys \citep{BrandtXRaySurveyResults2006}.
This difference could also be due to over-estimations of the bolometric AGN luminosity from \textsc{ProSpect}. 
We summarise these results in Table~\ref{tab:MethodSummary}.

\subsection{Emission Lines}

\begin{figure*}
    \centering
    \includegraphics[width = \linewidth]{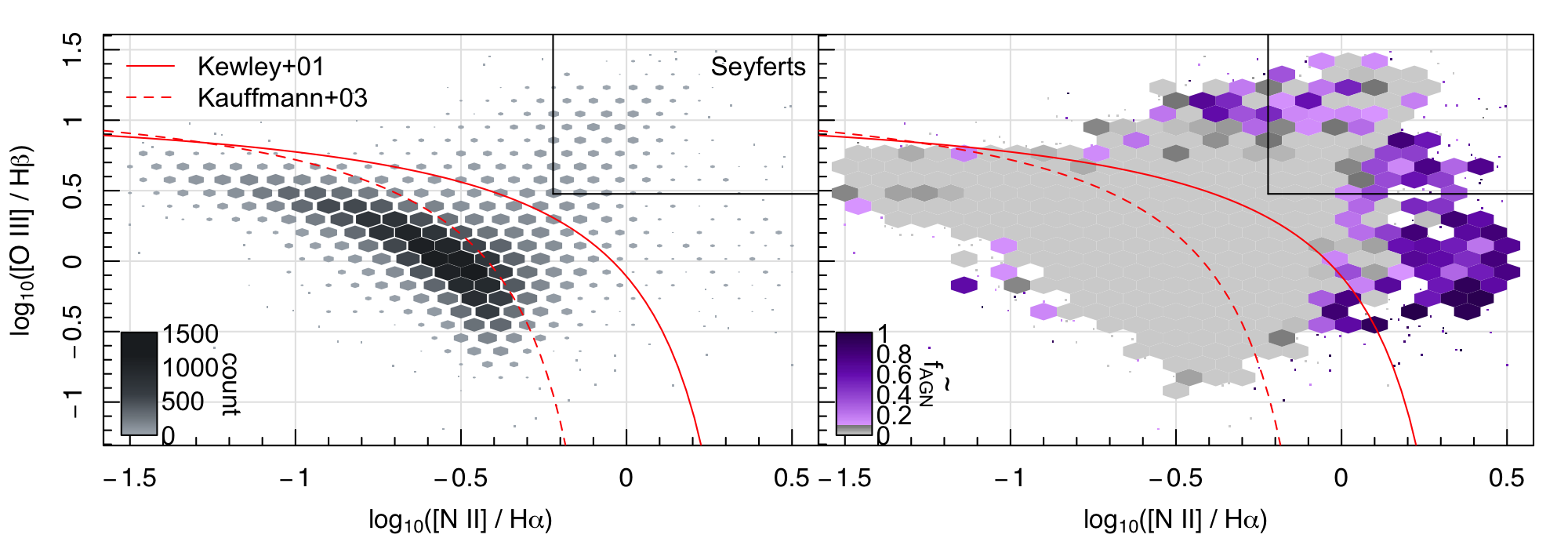}
    \caption{The distribution of the GAMA sample on the BPT \citep*{BaldwinClassificationparametersemissionline1981} diagram for objects with a S/N $>3$ for each of the four emission lines (left) and the parameter space coloured by median \textsc{ProSpect}-derived $f_\text{AGN}$ in each bin (right). 
    Bins with a median $f_\text{AGN} < 0.1$ are shown in grey, and bins with increasing $f_\text{AGN} > 0.1$ are shown in darker shades of purple. 
    The solid red curve shows the demarcation between starburst galaxies and AGN defined by \citet{KewleyTheoreticalModelingStarburst2001} while the dashed red curve shows the demarcation as defined by \citet{Kauffmannhostgalaxiesactive2003}.
    Using the prescription from \citet{Kauffmannhostgalaxiesactive2003} we define Seyfert objects as those with [O III]/H$\beta$ > 3 and  [N II]/H$\alpha$ > 0.6 (shown as the black lines).
    } 
    \label{fig:GAMABPT}
\end{figure*}

AGN can also be identified through the presence of broad and/or narrow emission lines in their optical spectra (e.g. \citealt{SeyfertNuclearEmissionSpiral1943,BaldwinClassificationparametersemissionline1981,KewleyTheoreticalModelingStarburst2001,MeyerQuasarscompletespectroscopic2001,Kauffmannhostgalaxiesactive2003}).
In this section we compare to a subset of galaxies from the DEVILS sample with spectra from the zCOSMOS survey \citep{LillyzCOSMOS10kBrightSpectroscopic2009} with broad line classifications and then compare the \textsc{ProSpect} and spectroscopic classifications for the GAMA sample.

\subsubsection{Broad Line AGN in zCOSMOS}

We use existing spectra from the zCOSMOS survey \citep{LillyzCOSMOS10kBrightSpectroscopic2009} that have already been classified and include 123 objects with a secure or very secure redshift and a broad line classification (confidence classes 13 and 14).
The majority of these objects are masked (\textit{mask} column) or counted as stars (\textit{starFlag}) in the \texttt{DEVILS\_PhotomCat} catalogue.
62 of these broad line objects are selected for fitting in this work but 9 of these objects do not have measured FIR photometry.
The redshift distribution of these sources is shown in Figure~\ref{fig:BLAGN}.
51 of the remaining 53 objects have a \textsc{ProSpect}-derived $f_\text{AGN} > 0.1$. 

The two objects that are not selected as AGN by \textsc{ProSpect} show very little visual evidence for a MIR excess. 
This agreement demonstrates that using \textsc{ProSpect} on multiwavelength photometry alone we are able to recover the AGN classifications that are derived through broad line spectra. 
\newline

\subsubsection{Narrow and Broad Lines in GAMA}\label{sec:NLBL_GAMA}

\begin{figure}
    \centering
    \includegraphics[width=\linewidth]{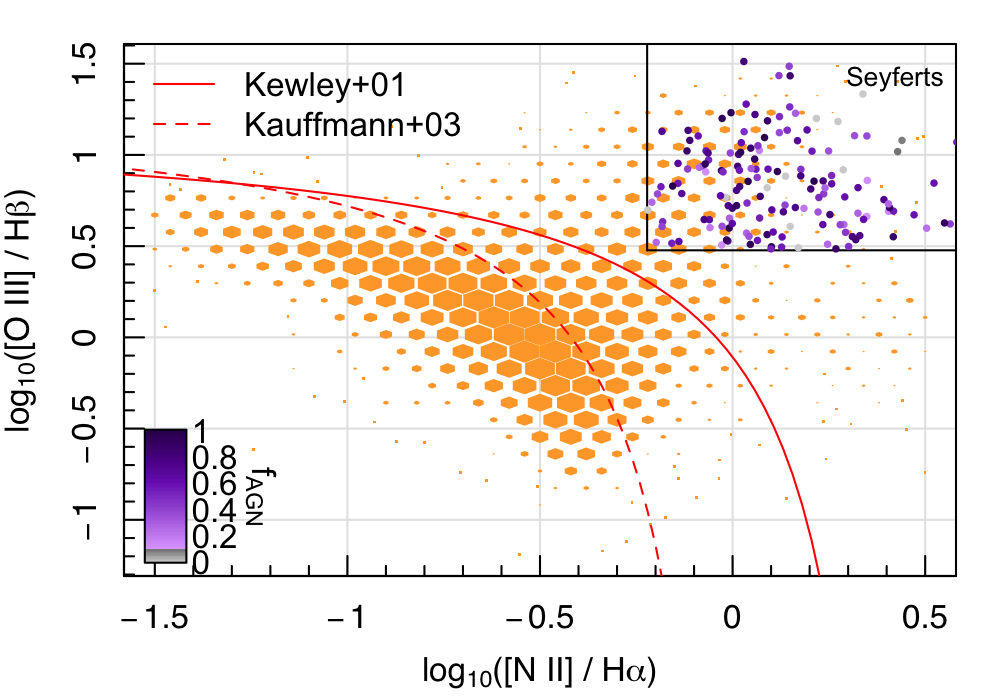}
    \caption{The BPT \citep{BaldwinClassificationparametersemissionline1981} diagram for the sample of galaxies from GAMA in which we plot the emission-line flux ratio [O III]/H$\beta$ versus the ratio [N II]/H$\alpha$.
    The orange 2D histogram shows the distribution of sources with $z<0.4$, where all four lines are detected with S/N > 3. 
    The solid red curve shows the demarcation between starburst galaxies and AGN defined by \citet{KewleyTheoreticalModelingStarburst2001} while the dashed red curve shows the demarcation as defined by \citet{Kauffmannhostgalaxiesactive2003}.
    Using the prescription from \citet{Kauffmannhostgalaxiesactive2003} we define Seyfert objects as those with [O III]/H$\beta$ > 3 and  [N II]/H$\alpha$ > 0.6 (shown as the black lines). 
    The sample of BPT and broad line selected AGN are shown as the coloured circles. 
    Objects with increasing values of $f_\text{AGN}$ are shown in darker shades and the colour transitions from grey scale to shades of purple at the $f_\text{AGN} = 0.1$ threshold.
    }
    \label{fig:G23_BPT}
\end{figure}

\begin{figure*}
    \centering
    \includegraphics[width = \linewidth]{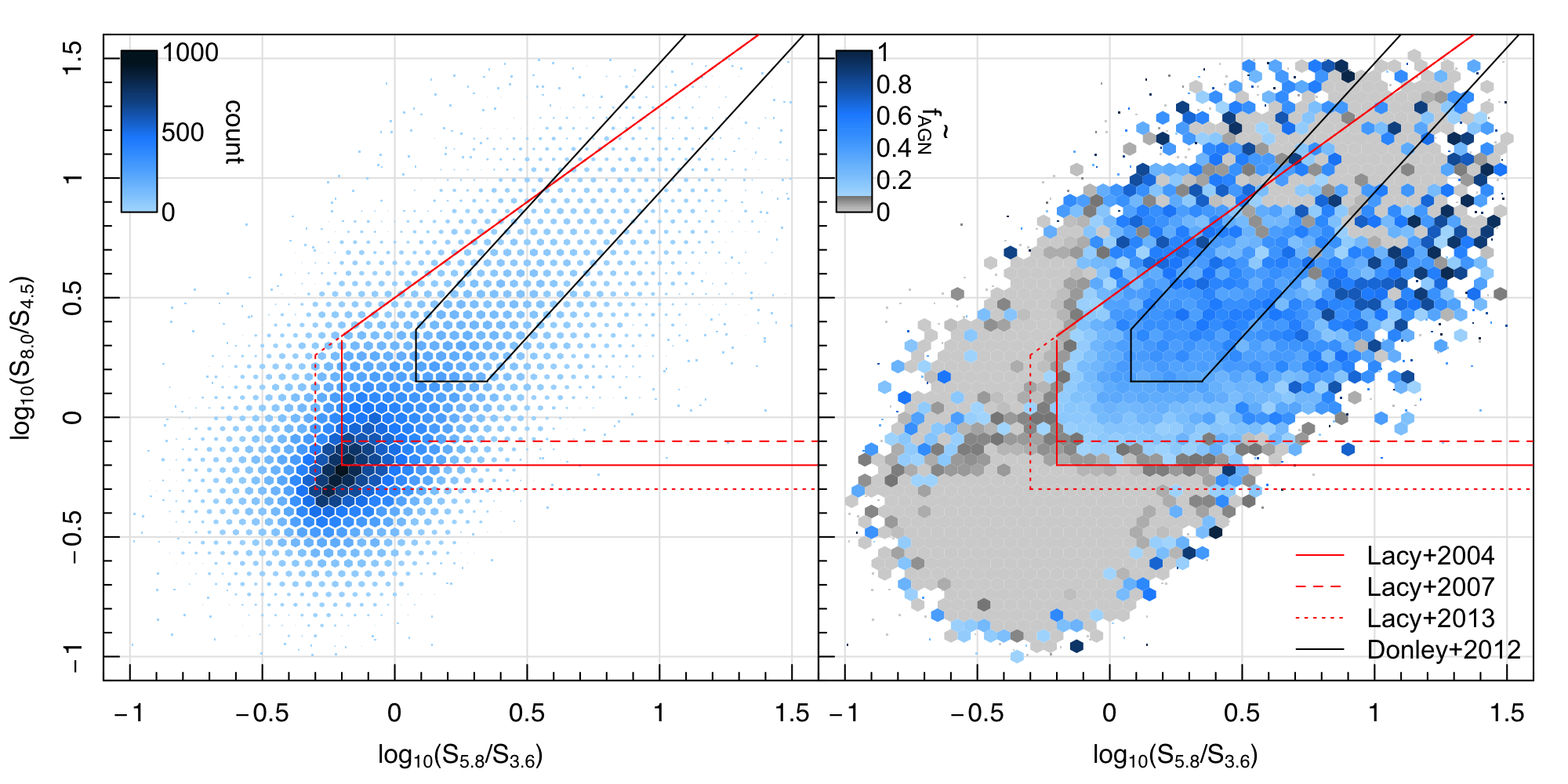}
    \caption{
    Comparison of Spitzer AGN selection criteria with \textsc{ProSpect} selected AGN for DEVILS objects with $\text{S/N}>5$ in each of the four IRAC channels. 
     The left panel shows the number distribution of the DEVILS sample in \textit{Spitzer} colour-colour space where both the colour and size of the hexagons represent the number of objects in each bin. 
     The right panel shows the \textit{Spitzer} colour-colour space coloured by median \textsc{ProSpect}-derived $f_\text{AGN}$ in each bin. 
    Increasing values of $f_\text{AGN} < 0.1$ are shown in greyscale, and increasing values $f_\text{AGN} > 0.1$ are shown in darker shades of blue. 
    We show the commonly used Spitzer colour selections from \citet{LacyObscuredUnobscuredActive2004,LacyOpticalSpectroscopyXRay2007,LacySpitzerMidinfraredActive2013} and \citet{DonleyIdentifyingLuminousActive2012} in both panels where AGN are selected from within the boundaries. 
    }
    \label{fig:SpitzerColours}
\end{figure*}

\begin{figure}
    \centering
    \includegraphics[width = \linewidth]{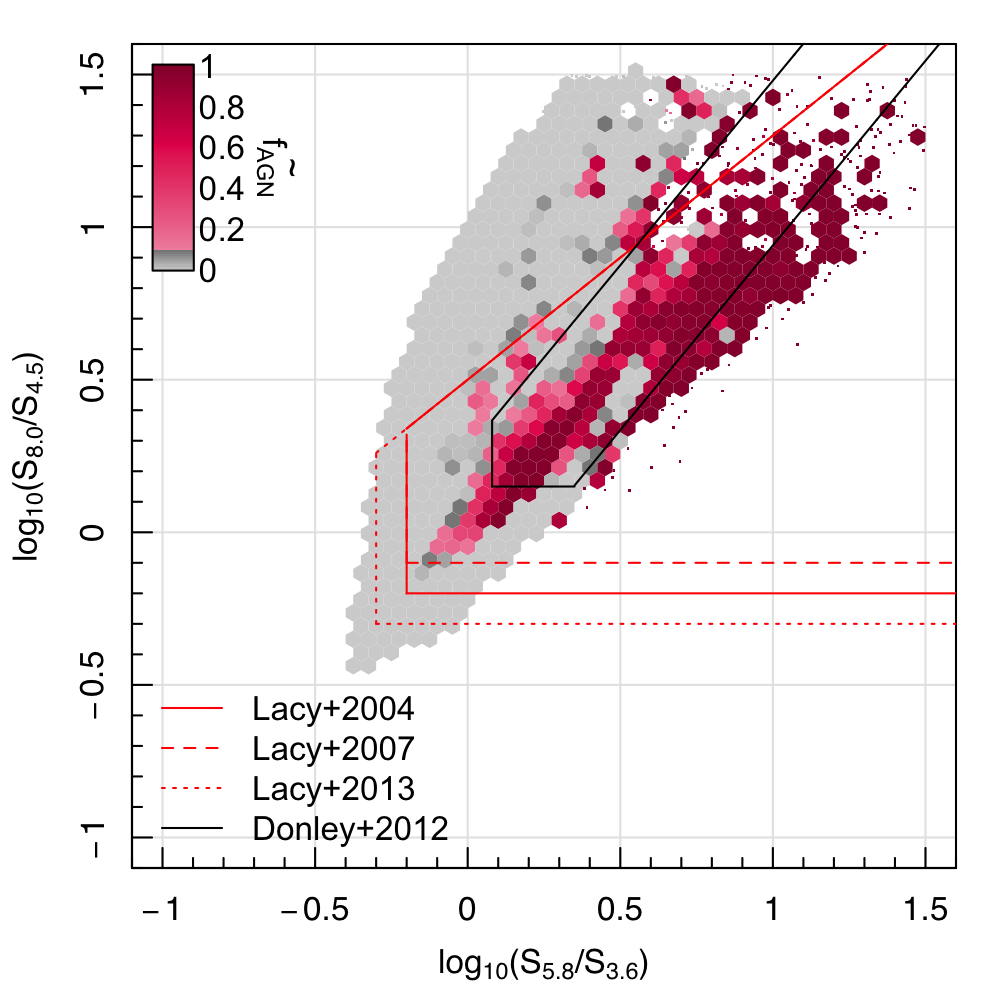}
    \caption{As per Figure~\ref{fig:SpitzerColours} but using 100,000 mock galaxies generated using \textsc{ProSpect} with $z < 3$ and parameters drawn from the ranges presented in Table~\ref{tab:Parameters}. 
    }
    \label{fig:SpitzerMock}
\end{figure}

One of the other AGN identification techniques is through narrow line flux ratios \citep{BaldwinClassificationparametersemissionline1981,KewleyTheoreticalModelingStarburst2001,Kauffmannhostgalaxiesactive2003,Kewleyhostgalaxiesclassification2006}.
As the DEVILS spectra do not have the signal-to-noise required to measure line fluxes, we turn to spectroscopic and photometric measurements from the GAMA survey.

To compare to emission line selected AGN in GAMA, we use the BPT diagram \citep*{BaldwinClassificationparametersemissionline1981} and the demarcations from \cite{KewleyTheoreticalModelingStarburst2001} and \cite{Kauffmannhostgalaxiesactive2003}.
We use the emission line measurements from the \texttt{SpecLinesGaussFitComplexv05} catalogue \citep{GordonGalaxyMassAssembly2017}.
For each object in GAMA, we use the best spectra (\texttt{IS\_BEST} = T) available regardless of survey as, although some are not flux calibrated, the flux ratios required for AGN identification are very close in wavelength are robust to poor calibration.
To ensure accurate line ratios for all objects we restrict our sample to objects with $z<0.4$ that have a signal-to-noise ratio $> 3$ in all four emission lines used in the BPT classification (i.e. H$\alpha$, H$\beta$, [N II] and [O III]).
Figure~\ref{fig:GAMABPT} shows the number distribution of the GAMA sources across the BPT diagram and the median $f_\text{AGN}$ in each bin. 
Bins with a median $f_\text{AGN} < 0.1$ are shown in grey and bins with increasing $f_\text{AGN} < 0.1$ are shown in darker shades of purple.
From Figure~\ref{fig:GAMABPT}, it is clear that \textsc{ProSpect} generally recovers objects with line ratios above the demarcation from \cite{KewleyTheoreticalModelingStarburst2001} and with high values of [N II]/H$\alpha$, typically associated with the lowest-luminosity AGN (low-ionization
nuclear emission regions; LINERs). 
There are however, a number of sources that would be identified as AGN using the \cite{KewleyTheoreticalModelingStarburst2001} demarcation, or as Seyferts that \textsc{ProSpect} does not recover.
This is not entirely unexpected as low accretion rate AGN (Seyferts, especially type 2s) are commonly associated with narrow line emission, but not significant MIR emission (e.g. \citealt{DeoMidInfraredContinuaSeyfert2009}).

BPT-selected samples can be contaminated with non-AGN as there is no clear separation between populations in the BPT parameter space, and because BPT-selected samples can be contaminated by poor measurements of emission line fluxes.
To select robust AGN from the BPT diagram we isolate a sample of AGN with narrow-line ratios indicative of an AGN and with clear broad lines to compare to \textsc{ProSpect}-selected AGN.

To first select narrow-line AGN, we use the Seyfert demarcation presented in figure 1 of  (\citealt{Kauffmannhostgalaxiesactive2003}; i.e. we require [O III]/H$\beta$ > 3 and [N II]/H$\alpha$ > 0.6).
As we are selecting sources with both broad and narrow lines present in the spectrum, we use the narrow component of the Balmer lines (H$\alpha$ and H$\beta$) included in the \texttt{SpecLinesComplexFitv05} catalogue \citep{GordonGalaxyMassAssembly2017}. 
These line measurements are obtained using multiple Gaussian components including a possible broad component for the hydrogen lines.
Using the narrow component ensures the line ratios do not include the broad component which would change the line ratios.

To ensure that the excitation in the line ratios is due to the presence of an AGN rather than another process (such as shocks or star formation)  we limit our selection to objects with broad H$\alpha$ or H$\beta$ emission. 
This will preferentially select for face-on AGN where the broad line regions are visible. 
We do this by first visually selecting objects with clear broad features in the spectra and then confirming through the width measurements in the GAMA \texttt{SpecLinesComplexFitv05}.

There are a number of galaxies in GAMA with a large angular size where the fibre probes only the emission from the very central region (i.e. possibly dominated by an AGN), while the SED could be dominated by the host galaxy. 
To ensure that the spectra and SED are measured over similar spatial scales we remove objects with an R50 > 4 arcsec.
The R50 value is measured by {\sc ProFound} during the photometry extraction and is the semi-major axis containing 50 per cent of the flux (see \citealt{BellstedtGalaxyMassAssembly2020a}).

This process of selecting AGN based on BPT line ratios, clear presence of broad lines, and small angular size results in a sample of 163 galaxies. 
\textsc{ProSpect} recovers an $f_\text{AGN} > 0.1$ for 149/163 of these galaxies (91.4 per cent). 
Figure~\ref{fig:G23_BPT} shows the BPT distribution of all sources in the GAMA catalogue with $z<0.4$ and S/N > 3 in each of the four emission lines. 
We show the positions of the 163 objects on the BPT as the coloured circles, where objects with \textsc{ProSpect}-derived $f_\text{AGN} > 0.1$ are shown in blue and $f_\text{AGN} < 0.1$ shown in grey.
The objects that are not selected as AGN by \textsc{ProSpect} have high stellar masses $M_\star > 10^{10.5}M_\odot$, and lie on or above the star-forming main sequence (see \citealt{ThorneDeepExtragalacticVIsible2021}).
The presence of an AGN in the SED of these galaxies could be concealed by the high luminosity of the host galaxy.
Visually inspecting the SEDs and \textsc{ProSpect} fits we find no evidence of a MIR excess for any of these galaxies.

\subsection{MIR Colours}

One of the other commonly used methods to identify AGN is through MIR colours (i.e. \citealt{LacyObscuredUnobscuredActive2004, LacyOpticalSpectroscopyXRay2007,AssefLowResolutionSpectralTemplates2010,JarrettSPITZERWISESURVEY2011,SternMIDINFRAREDSELECTIONACTIVE2012,AssefMidInfraredSelectionActive2013,LacySpitzerMidinfraredActive2013}). 
Out of all the different AGN selection techniques discussed here, SED selection is most similar to MIR colour selections as the wavelength range that drives the constraint of an AGN component in SED fitting is the MIR. 
We expect the greatest overlap between \textsc{ProSpect} and MIR colour selections as the value of $f_\text{AGN}$ is also derived from emission in the MIR.
Although MIR colour selections are simple and can, in some cases, be implemented with measurements in only two photometric bands, they can be contaminated by other populations. 
The MIR selection techniques rely on the typically redder colours of AGN, particularly between 3-5\,$\mu$m. \
However, there are a number of other populations that can mimic the colours of AGN in these bands and contaminate the sample.
This includes star forming galaxies at $z\sim0.2$ with powerful polycyclic aromatic hydrocarbon emission, or massive galaxies at high redshift ($z>1$) for which the stellar bump is shifted into the MIR. 
SED fitting allows us to disentangle the contribution from the host galaxy, quantify the luminosity of the AGN, and also allow for the shift of the observed SED to longer wavelengths with increasing redshift. 

In the following sections we use compare various MIR selection criteria to AGN identified using \textsc{ProSpect}.
Due to the different photometric coverage between surveys, we use DEVILS to compare to selections derived using \textit{Spitzer}, and GAMA to compare to WISE-based selection criteria.

\begin{figure*}
    \centering
    \includegraphics[width = \linewidth]{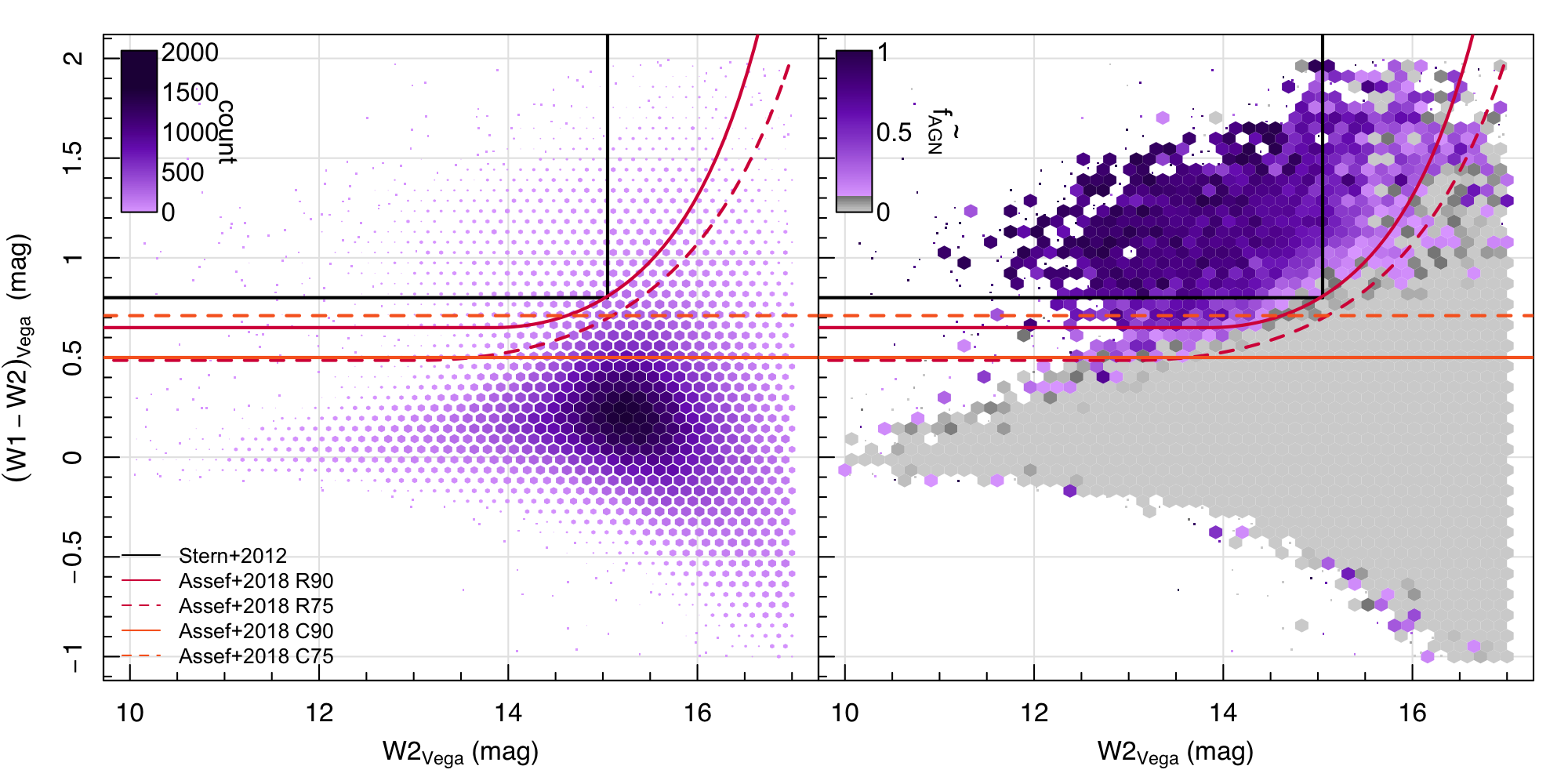}
    \caption{
    The left panel shows the number distribution of the GAMA sample in WISE colour-magnitude space where both the colour and size of the hexagons represent the number of objects in each bin. 
    The right panel shows the WISE colour-magnitude coloured by median \textsc{ProSpect}-derived $f_\text{AGN}$ in each bin. 
    Bins with a median $f_\text{AGN} < 0.1$ are shown in grey, and bins with increasing $f_\text{AGN} > 0.1$ are shown in darker shades of purple. 
    We show the commonly used WISE colour selections from \citet{SternMIDINFRAREDSELECTIONACTIVE2012} and \citet{AssefWISEAGNCatalog2018} in both panels. The \citet{AssefWISEAGNCatalog2018} relations are shown for the 75 (dashed) and 90 per cent (solid) completeness (orange) and reliability (red) cuts. }
    \label{fig:WISE}
\end{figure*}

\subsubsection{Spitzer Colours}\label{sec:spitzer}

An IRAC colour-colour diagram, using all four broad band channels of the IRAC instrument is shown in Figure~\ref{fig:SpitzerColours}.
The left panel shows the number distribution of objects from the DEVILS sample and a signal-to-noise ratio, S/N$>5$ in each of the four bands.
In this panel both the colour and size of the hexagons represent the number of galaxies in each bin. 
We also show this parameter space coloured by median $f_\text{AGN}$ derived by \textsc{ProSpect} to show a comparison to the MIR selection criteria presented in \citet{LacyObscuredUnobscuredActive2004,LacyOpticalSpectroscopyXRay2007,LacySpitzerMidinfraredActive2013} and \citet{DonleyIdentifyingLuminousActive2012}.
We find that in general objects with $f_\text{AGN} < 0.1$ are also selected by these MIR criteria, but there are a number of bins with median $f_\text{AGN} >0.1$ outside the \citet{LacyObscuredUnobscuredActive2004,LacyOpticalSpectroscopyXRay2007,DonleyIdentifyingLuminousActive2012,LacySpitzerMidinfraredActive2013} AGN region. 
These are often due to poor photometry in a single band incorrectly biasing the flux ratio. 
AGN with confused photometry would be missed by the simple colour-colour selections. 
By incorporating the rest of the SED, especially measurements in the NIR and FIR, these issues can be mitigated and a more complete selection of AGN can be obtained.

The \textit{Spitzer} AGN selection criteria presented in \citet{LacyObscuredUnobscuredActive2004,LacyOpticalSpectroscopyXRay2007,LacySpitzerMidinfraredActive2013} are designed to have high completeness, but are not optimised for reliability. 
Of objects selected as AGN using the criterion from \cite{LacyObscuredUnobscuredActive2004}, we recover 54 per cent of these as AGN with \textsc{ProSpect} (i.e. $f_\text{AGN} > 0.1$).
To demonstrate this we randomly generate 100,000 SEDs with \textsc{ProSpect} using the ranges for each parameter presented in Table~\ref{tab:Parameters}.
We show the recovered IRAC colour-colour diagram in Figure~\ref{fig:SpitzerMock} coloured by $f_\text{AGN}$.
Using generated photometry rather than measured photometry shows the possible colour combinations from the model for objects with a significant AGN component without contamination from poor measurements. 
It is clear from Figure~\ref{fig:SpitzerMock} that the \citet{LacyObscuredUnobscuredActive2004,LacyOpticalSpectroscopyXRay2007,LacySpitzerMidinfraredActive2013} AGN selection criteria recover almost all AGN colour combinations generated by the \cite{FritzRevisitinginfraredspectra2006} and \cite{FeltreSmoothclumpydust2012} model.
However it is clear that there is significant contamination in this selection from objects with no AGN component, demonstrating the low reliability of these criteria. 

Although the \cite{DonleyIdentifyingLuminousActive2012} criteria are designed for higher reliability than those from \citet{LacyObscuredUnobscuredActive2004,LacyOpticalSpectroscopyXRay2007,LacySpitzerMidinfraredActive2013}, it is apparent from Figure~\ref{fig:SpitzerMock} that there are a significant number of AGN colour combinations that lie outside the criteria, reducing the completeness of these criteria. 
There are also colour combinations selected as AGN by the \citet{DonleyIdentifyingLuminousActive2012} criteria for which we find no significant AGN contribution on average.

\subsubsection{WISE Colours}
To compare our SED-selected AGN to commonly-used WISE colour selections we use the fits to the sub-sample of GAMA galaxies with a signal-to-noise (S/N) ratio $>5$ in \textit{W1} and \textit{W2}.
We use the entire redshift range covered by these objects but note that the vast majority of objects in GAMA have $z<0.4$.
This results in a sample size of 232,076.

Figure~\ref{fig:WISE} shows the distribution of the GAMA sample in colour (W1-W2) versus magnitude (W2) space and the parameter space coloured by median $f_\text{AGN}$.
We also show commonly used AGN selection criteria from \cite{SternMIDINFRAREDSELECTIONACTIVE2012} and  \cite{AssefWISEAGNCatalog2018} where AGN are selected if they exist in the upper left corner. 
As expected due to the nature of the selection techniques, it is clear from this comparison that there is good agreement between \textsc{ProSpect} and WISE colour selected AGN.
However, there are sources selected as AGN through the WISE colour cuts that are not selected using \textsc{ProSpect} and vice versa.
For example using the \cite{SternMIDINFRAREDSELECTIONACTIVE2012} selection (shown as the black rectangle), \textsc{ProSpect} recovers 81.8 per cent of these objects as AGN. 
This number drops to 70.7 per cent using the \cite{AssefWISEAGNCatalog2018} 90 per cent reliability cut.
\textsc{ProSpect} is also able to recover AGN that are fainter in W2 than the reliability cuts from \cite{AssefWISEAGNCatalog2018}.
We note that \textsc{ProSpect} does miss objects that would be selected as AGN using the \cite{AssefWISEAGNCatalog2018} completeness selections, but this is in the region with low reliability.
As with the \textit{Spitzer} colour selections, poor photometry measurements in either band can bias the colour measurement for AGN, incorrectly placing them outside the AGN selection region.
\newline

In summary, by combining the results from \textsc{ProSpect} with previous multiwavelength AGN diagnostics we demonstrate that \textsc{ProSpect} recovers 54.3 and 69.0 per cent of AGN selected through the \citet{LacySpitzerMidinfraredActive2013} and \citet{DonleyIdentifyingLuminousActive2012} Spitzer criteria respectively. 
\textsc{ProSpect} also recovers between 42-82 per cent of AGN selected by the WISE criteria defined in \citet{SternMidInfraredSelectionActive2005} and \citet{AssefWISEAGNCatalog2018} depending on the reliability or completeness selections used. 
We also demonstrate, using the GAMA sample, that \textsc{ProSpect} recovers 38.9 and 21.3 per cent of AGN selected using the BPT selections from \citet{KewleyTheoreticalModelingStarburst2001} and \citet{Kauffmannhostgalaxiesactive2003}.
We find that \textsc{ProSpect} recovers an AGN component in 91.4 per cent of a subsample of Seyferts with visible broad lines (Seyfert 1s) from the GAMA survey. 
Finally, by cross-matching to the \textit{Chandra}-COSMOS Legacy catalogue described by \citet{MarchesiCHANDRACOSMOSLEGACY2016}, we demonstrate that \textsc{ProSpect} recovers and AGN component in 65 per cent of X-ray detected sources. 
However, we find that the \textsc{ProSpect}-derived AGN luminosities predict far more X-ray sources above the \textit{Chandra} sensitivity limit than detected.
Each of these comparisons are summarised in Table~\ref{tab:MethodSummary}.

\section{Comparisons to other AGN quantification methods}\label{sec:AGNQuantification}

While we have shown that \textsc{ProSpect} can identify AGN that are also selected by MIR, emission line and X-ray techniques, \textsc{ProSpect} can also be used simultaneously to quantify the emission from an AGN component. 
In the following sections we compare the \textsc{ProSpect}-derived AGN luminosities to those derived from X-ray emission and from another SED-fitting code.

\subsection{X-ray}
As both the X-ray and MIR emission from AGN are related to the accretion rate of the SMBH, AGN luminosities measured from both regimes should be correlated.
Differences can arise due to X-ray emission depending on properties of the hot corona gas density, temperature etc and the MIR emission depending on geometry. 
If \textsc{ProSpect} has accurately detected and quantified AGN emission in these galaxies we expect good agreement with AGN luminosities measured from X-ray imaging. 
We stress that \textsc{ProSpect} does not use X-ray measurements in fitting the SEDs, and agreement with X-ray measurements is not a guaranteed outcome of the modelling. 

Using the position-matched catalogue described in Section~\ref{sec:xray}, we compare the luminosities obtained from the \textit{Chandra}-COSMOS legacy survey to those obtained by \textsc{ProSpect}.
We limit our comparisons to objects with an X-ray counterpart and a  \textsc{ProSpect}-derived $f_\text{AGN} > 0.1$ (i.e. both \textsc{ProSpect} and X-ray selected AGN), resulting in a sample of 1,528 AGN.

\begin{figure}
    \centering
    \includegraphics[width = \linewidth]{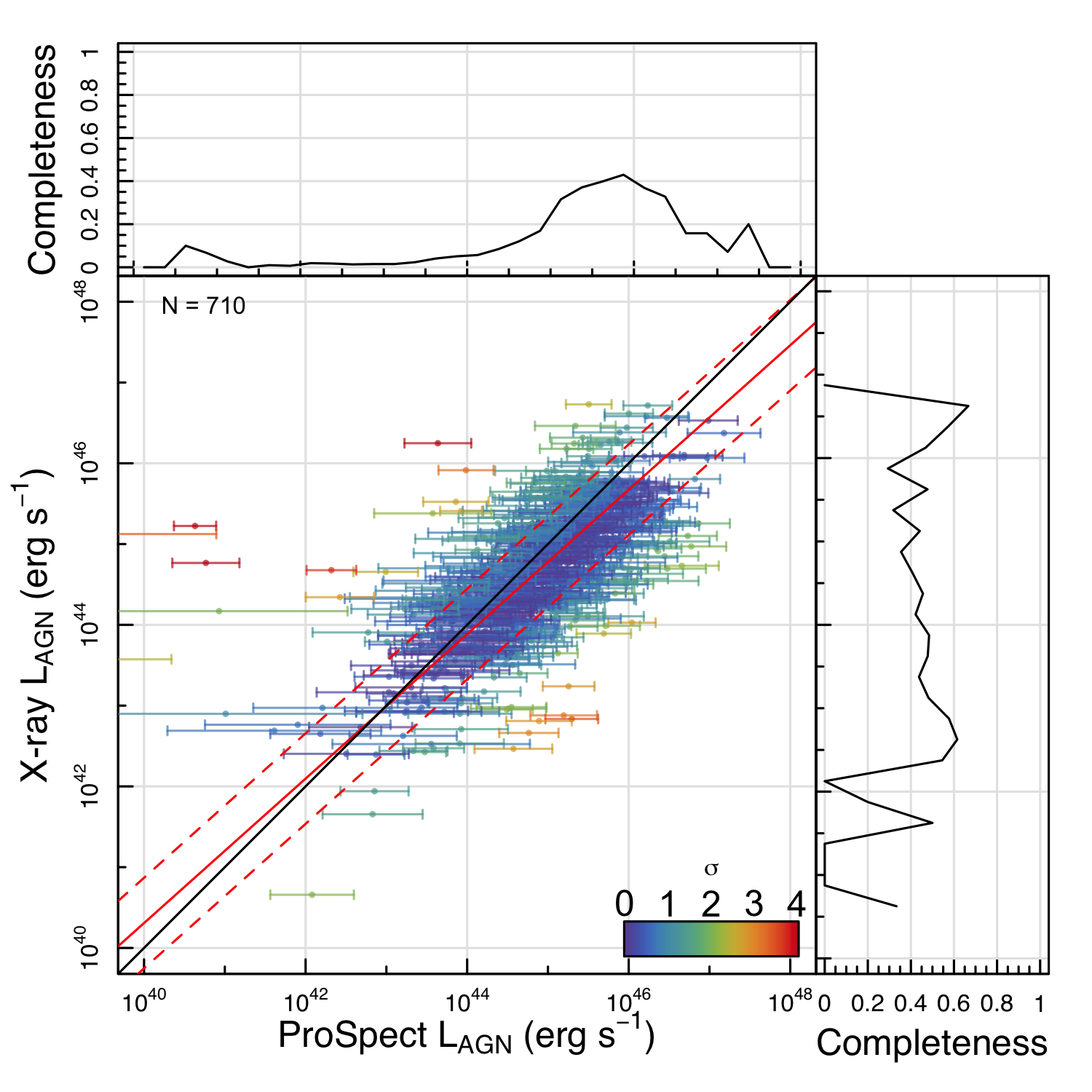}
    \caption{The bolometric AGN luminosity derived from \textsc{ProSpect} compared to the bolometric luminosity derived from the hard X-ray luminosities from \textit{Chandra} for objects with a \textsc{ProSpect}-derived $f_\text{AGN} > 0.1$.
    The solid red line shows the best linear fit from \textsc{Hyper-fit} and the dashed lines show the one standard deviation range. 
    The points are coloured by their `sigma-tension' from the linear fit and the black line denotes the one-to-one relation. 
    The top panel shows the fraction of objects that are \textsc{ProSpect}-selected AGN and detected by \textit{Chandra} out of those with predicted fluxes above the \textit{Chandra} sensitivity limit.
    The right panel shows the fraction of \textit{Chandra} selected AGN that are also selected as AGN by \textsc{ProSpect} as a function of X-ray AGN luminosity. 
    }
    \label{fig:XrayComp}
\end{figure}

In Figure~\ref{fig:XrayComp}, we show the comparison between our derived bolometric AGN luminosities from \textsc{ProSpect} to those from X-ray measurements.
We convert the \textit{Chandra} hard X-ray luminosities (2-10keV) to bolometric AGN luminosities using the correction from \cite{NetzerBolometriccorrectionfactors2019} as follows:
\begin{equation} \label{eq:bolcorrect}
    L_\textsc{BOL} = 7 [L_{\rm{2-10keV}} / (10^{42} \text{ erg sec}^{-1})]^{0.3} \times L_{\rm{2-10keV}}.
\end{equation}

We use \textsc{Hyper-fit} \citep{RobothamHyperFitFittingLinear2015} to fit the comparison and recover a slope of $0.807 \pm 0.03 $, with an orthogonal scatter of $0.46 \pm 0.01$\,dex. 
This implies good correlation between the \textsc{ProSpect}-derived AGN luminosities and the corrected X-ray luminosities from \textit{Chandra}.

As mentioned above, there are X-ray detected AGN that are not identified as AGN using \textsc{ProSpect} and vice versa.
To calculate the number of \textsc{ProSpect}-selected AGN that we expect to be detected by \textit{Chandra}, we convert the \textsc{ProSpect}-derived AGN luminosity to a hard X-ray flux density using Equation~\ref{eq:bolcorrect}. 
Objects with a predicted hard X-ray flux density above $7.3 \times 10^{-16} \text{\,erg\,cm}^{-2}\text{\,s}^{-1}$ are above the sensitivity limit of \textit{Chandra} and should be detected.
The top panel of Figure~\ref{fig:XrayComp} shows the fraction of detected objects at each luminosity out of those with predicted fluxes above the \textit{Chandra} sensitivity limit. 
This fraction peaks at 40 per cent for objects with $L_\text{AGN} \approx 10^{46}$\,erg s$^{-1}$. 
We also show the fraction of X-ray selected AGN also detected as AGN from the SED in the right panel. 
This fraction is generally higher than the top panel and peaks at 70 per cent for objects with $L_\text{AGN} \approx 10^{47}$\,erg s$^{-1}$. 
This shows that \textsc{ProSpect} generally recovers the presence of an AGN in 50 per cent of X-ray selected objects across all luminosities. 
However, \textsc{ProSpect} recovers far more AGN than detected in X-rays. 

\begin{figure}
    \centering
    \includegraphics[width = \linewidth]{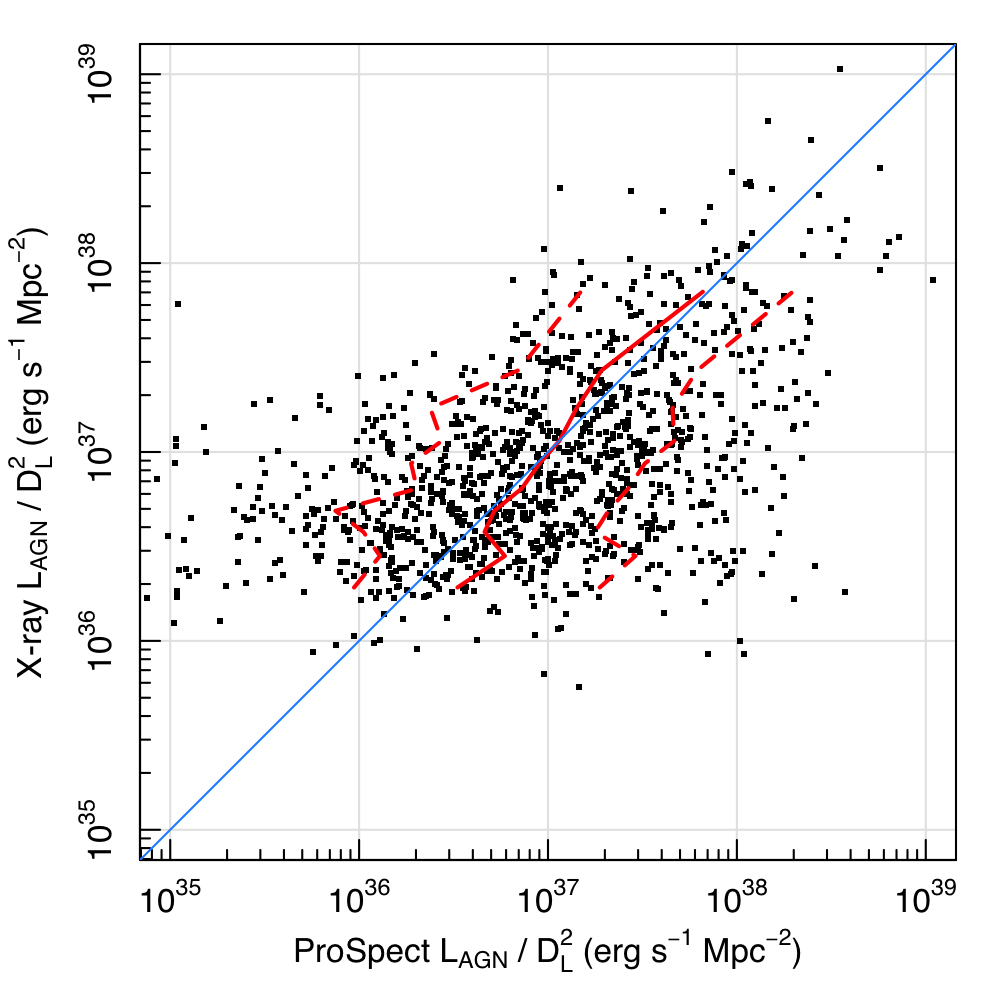}
    \caption{As per Figure~\ref{fig:XrayComp} but showing the luminosities divided by the square of the luminosity distance $(D_L^2)$ to remove the implicit distance bias when comparing luminosities. 
    We show the running median binned in \textit{Chandra}-derived luminosities (red) and the one-sigma range (dashed red).
    We also show the one-to-one line in blue. }
    \label{fig:Xraypseudoflux}
\end{figure}

When comparing luminosities directly, the correlation can be strengthened by the implicit inclusion of distance in the luminosity value. 
To demonstrate that the correlation we find between \textsc{ProSpect} and X-ray derived bolometric luminosities is not driven purely by distance we show the luminosities divided by the square of the luminosity distance ($D_L^2$) in Figure~\ref{fig:Xraypseudoflux}. 
The running median shows that \textsc{ProSpect} and X-ray derived bolometric luminosities still follow a one-to-one trend when accounting for distance, albeit with more scatter. 

\begin{figure}
    \centering
    \includegraphics[width = \linewidth]{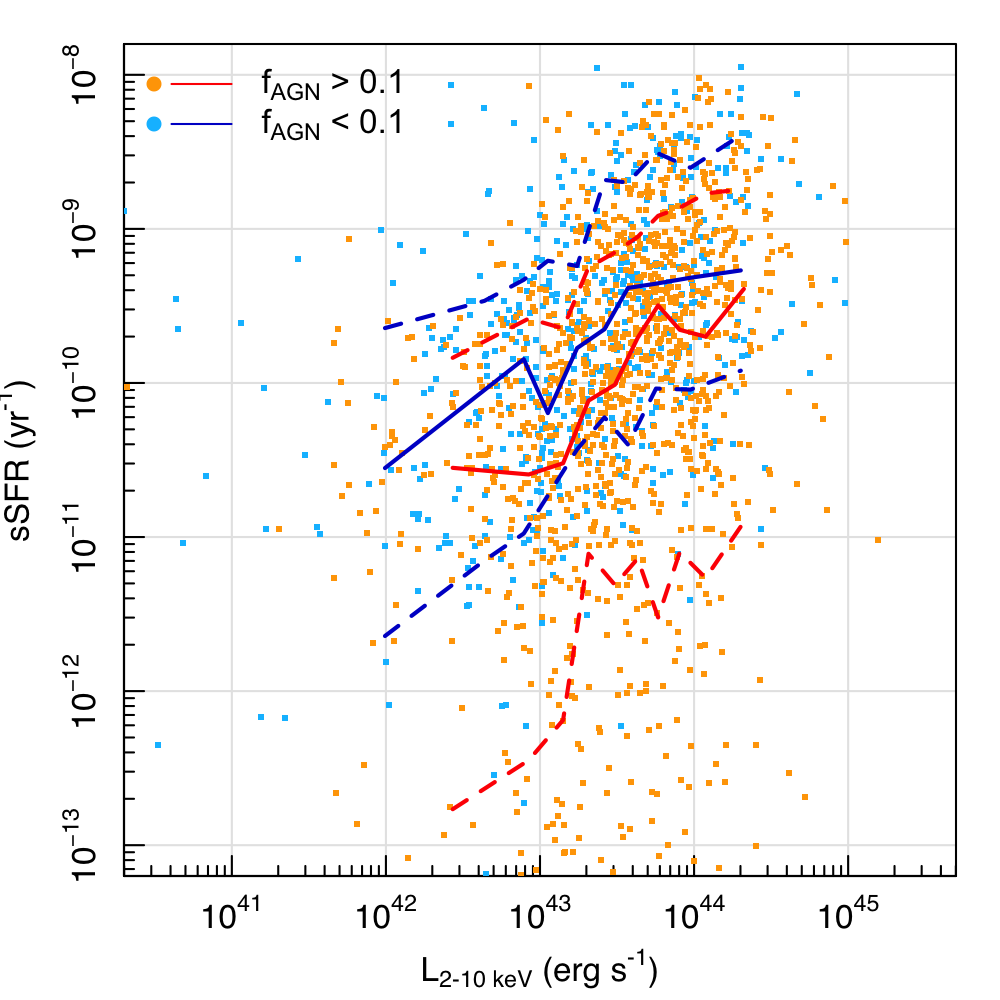}
    \caption{\textsc{ProSpect}-derived specific SFRs as a function of hard X-ray luminosity for all objects with a counterpart in the \textit{Chandra} COSMOS-legacy catalogue presented in \citet{MarchesiCHANDRACOSMOSLEGACY2016}. 
    We show all objects with $f_\text{AGN} < 0.1$ in light blue with the running median and one standard deviation ranges shown in dark blue and objects with $f_\text{AGN} > 0.1$ in orange points and red lines respectively. 
    }
    \label{fig:XraySFR}
\end{figure}

Figure~\ref{fig:XraySFR} shows the \textsc{ProSpect}-derived specific SFR (sSFR) and hard X-ray luminosity of all objects from this work with a counterpart in the \textit{Chandra} COSMOS-legacy catalogue.
At a given luminosity objects with $f_\text{AGN} < 0.1$ have higher sSFRs than objects with significant AGN contribution.
This offset could be for two reasons.
Firstly, for objects with $f_\text{AGN} < 0.1$, \textsc{ProSpect} could be incorrectly attributing elevated emission across the UV-optical regime with increased star-formation instead of a contribution from an AGN accretion disc. 
Or, the X-ray emission of objects with $f_\text{AGN} < 0.1$ could originate from stellar processes rather than from an AGN component in which case we would not expect an AGN to be identified from the MIR emission.
Distinguishing between these possibilities is beyond the scope of this work.

\subsection{Spectral Energy Distributions}
\begin{figure}
    \centering
    \includegraphics[width = \linewidth]{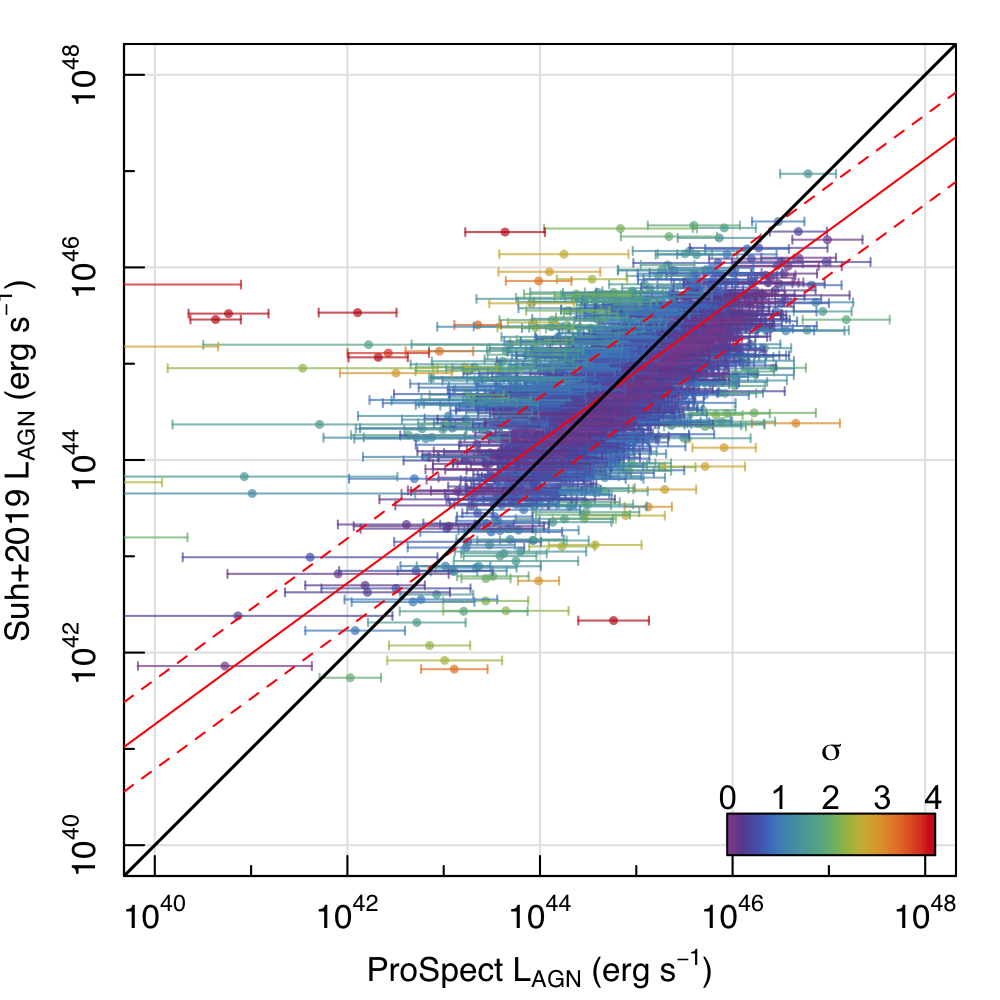}
    \caption{The bolometric AGN luminosity derived from \textsc{ProSpect} compared to that derived in \citet{SuhMultiwavelengthPropertiesType2019}. 
    We show objects with a \textsc{ProSpect}-derived $f_\text{AGN} > 0.1$.
    The solid red line shows the best linear fit from \textsc{Hyper-fit} and the dashed lines show the one standard deviation range. 
    The points are coloured by their `sigma-tension' from the linear fit and the black line denotes a one-to-one relation. 
    }
    \label{fig:SEDAGNLum}
\end{figure}

With the prior knowledge of the presence of an AGN within a galaxy, SED fitting tools have, in recent years, been applied to characterise these AGN \citep{CalistroRiveraAGNfitterBayesianMCMC2016,YangxcigalefittingAGN2020}. 
While we are not limited to `known' AGN, as our approach is blind, in the sense that we do not rely on prior information of AGN to characterise them, we asses how similar our results are to those that are extracted from other such techniques. 

\cite{SuhMultiwavelengthPropertiesType2019} derived AGN luminosities of X-ray selected AGN from \textit{Chandra} using a similar SED fitting code to \textsc{AGNfitter} \citep{CalistroRiveraAGNfitterBayesianMCMC2016}.
They use predetermined AGN types from optical properties to class AGN as Type 1 (Type 2), if they have broad lines (no broad lines) and/or their photometric redshift measurements were obtained using an unobscured (obscured) AGN template.
Both types of AGN were fit with a nuclear torus, a host galaxy, and a starburst component while the Type 1 AGN were fit with an additional big blue bump component in the UV-optical range. 
The host galaxy was modelled using a simple exponentially declining star formation history and fixed constant solar metallicity. 
We direct the reader to \citealt{CarnallHowMeasureGalaxy2019,LejaHowMeasureGalaxy2019,LowerHowWellCan2020} for a discussion on the impacts of the choice of star formation history on galaxy properties and \citealt{BellstedtGalaxyMassAssembly2020b,ThorneDeepExtragalacticVIsible2021} for a discussion on the impacts of a poorly motivated chemical enrichment history. 

Figure~\ref{fig:SEDAGNLum} shows the comparison between the \textsc{ProSpect} derived AGN luminosities and those derived by \cite{SuhMultiwavelengthPropertiesType2019} for objects with a \textsc{ProSpect} $f_\text{AGN} > 0.1$. 
As the objects fit by \cite{SuhMultiwavelengthPropertiesType2019} were selected from the \textit{Chandra} sample shown in Section~\ref{sec:xray}, this sample will have the same completeness fractions as shown in Figure~\ref{fig:XrayComp}.
Using \textsc{Hyper-fit} we derive a linear fit with a slope of $0.733 \pm 0.0005 $ and an orthogonal scatter of $0.37 \pm 0.0001\,$dex. 
As such we find close agreement with the AGN luminosities derived by \cite{SuhMultiwavelengthPropertiesType2019} despite the differences in SED modelling choices.
When comparing to the stellar masses derived by \cite{SuhMultiwavelengthPropertiesType2019} we recover stellar masses $\sim$0.2\,dex higher due to the implementation of the star formation and metallicity history in \textsc{ProSpect} which recovers older stellar populations with higher mass-to-light ratios and therefore more massive galaxies. 
While codes such as \textsc{AGNfitter} and the variant used by \cite{SuhMultiwavelengthPropertiesType2019} are specifically constructed to recover AGN properties while simplifying the star formation history, metallicity and dust of the host galaxy, \textsc{ProSpect} recovers consistent AGN luminosities while also allowing for more flexibility in the star formation and metallicity history of the host galaxy.

\section{The Impact of an AGN component on Derived Galaxy Properties}\label{sec:EffectOnHostGalaxies}
\begin{figure*}
    \centering
    \includegraphics[width = 0.48\linewidth]{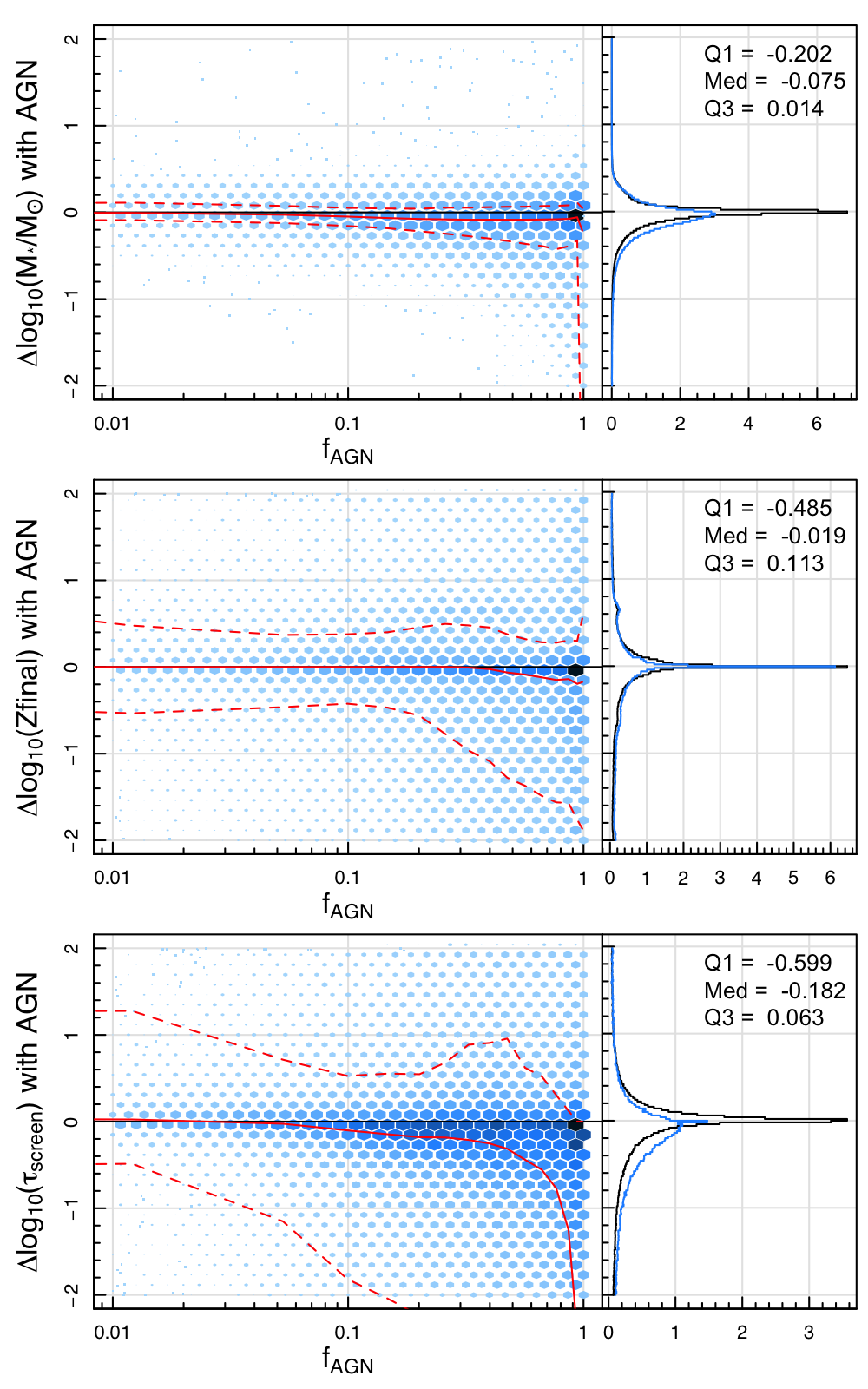}
    \hspace{3mm}
    \includegraphics[width = 0.48\linewidth]{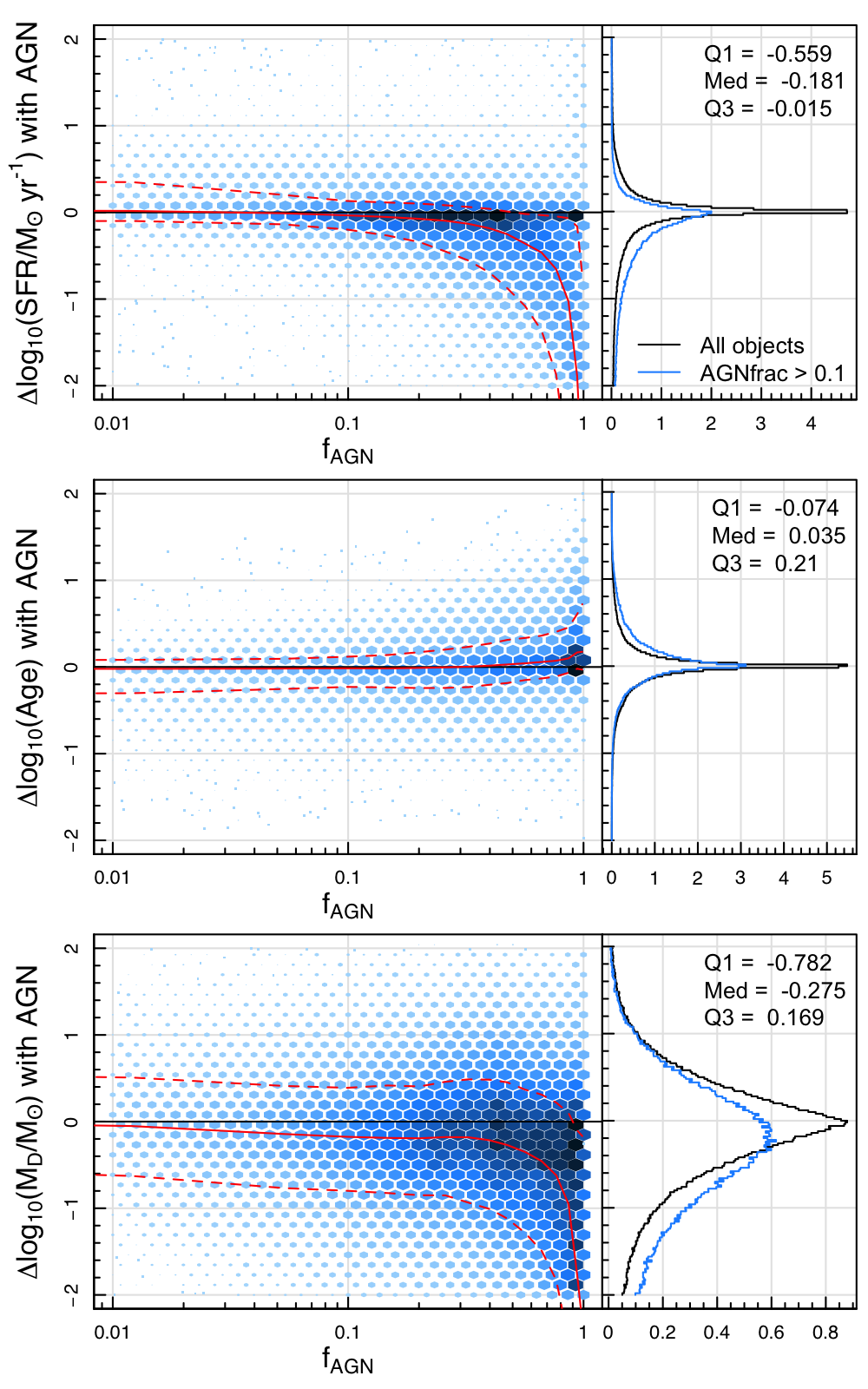}
    \caption{The binned change in galaxy properties as a function of $f_\text{AGN}$ between fits with and without AGN for objects with $f_\text{AGN} > 0.01$ shown as a 2D histogram. Both the colour and size of the hexagons shows the number of sources in each bin. 
    The running median is shown as the solid line with the running one sigma range shown as the dashed red lines. 
    We show the normalised distribution of change in each parameter on the right with the distribution of the whole sample (black) and the normalised distribution of the sample with $f_\text{AGN} > 0.1$ (blue). 
    The median, upper and lower quartiles of the change for the AGN sample are shown in the upper right corner. 
    The dust mass shown here refers only to the birth cloud and general ISM dust and does not include the mass of dust in the AGN torus which should be negligible. 
    }
    \label{fig:GalChanges}
\end{figure*}

\begin{figure*}
    \centering
    \includegraphics[width = \linewidth]{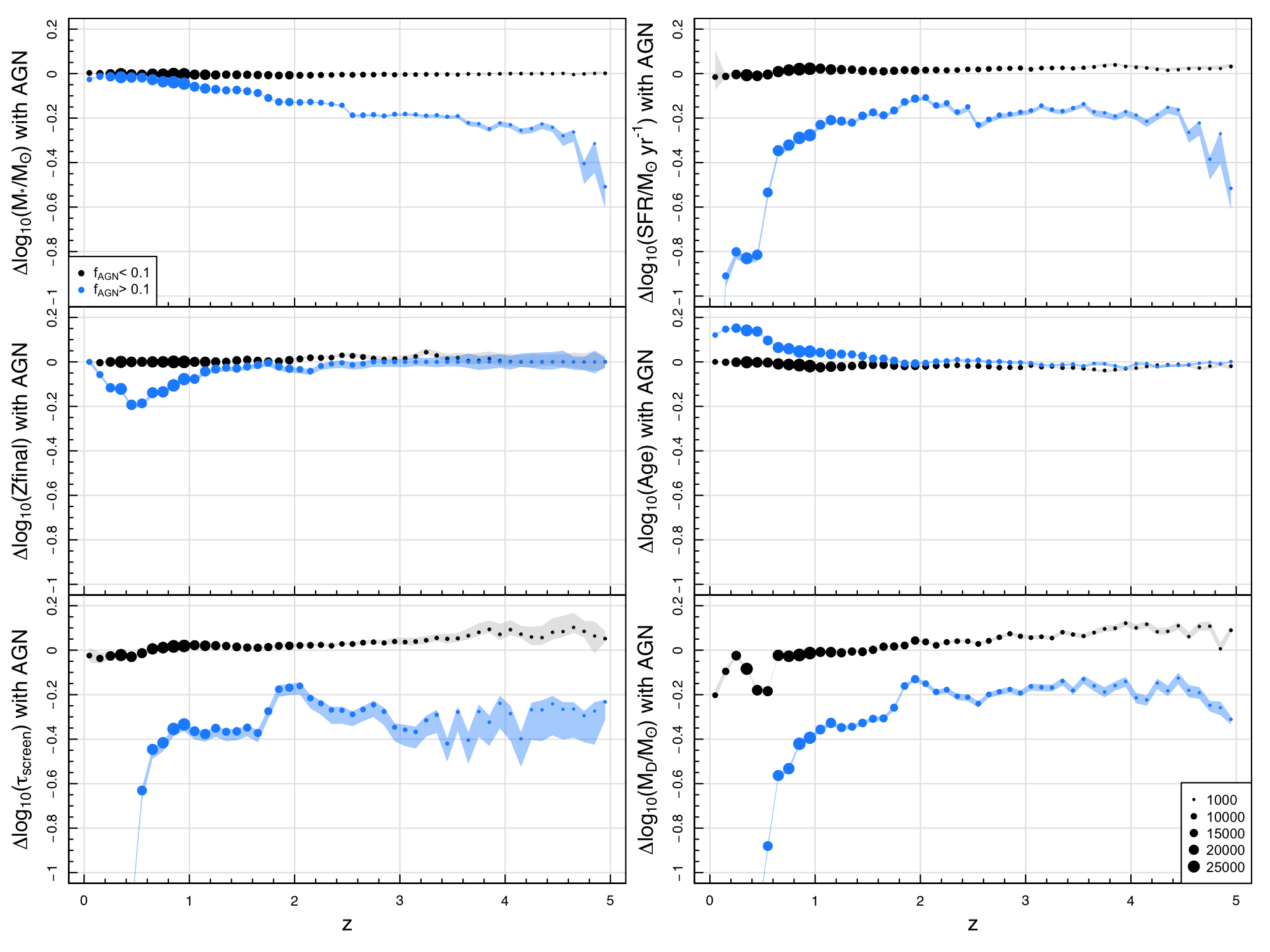}
    \caption{Median offset in parameters between fits with and without an AGN component from Figure~\ref{fig:GalChanges} as a function of redshift. 
    We show objects with $f_\text{AGN} > 0.1$ in blue (i.e. significant AGN component) and objects with $f_\text{AGN} < 0.1$ in black. 
    The shaded areas of the same colours show the standard error of the median and the point size corresponds to the number of objects in that redshift bin. }
    \label{fig:GalChangesWithz}
\end{figure*}

As the contribution from an AGN can span the entire wavelength range covered by typical SED modelling (FUV-FIR), the inclusion of an AGN component can have a significant impact on the recovered galaxy properties. 
\cite{LejaHotDustPanchromatic2018} find that the inclusion of an AGN in broad band SED fitting can change stellar ages and SFRs by up to an order of magnitude, and dust attenuation by up to a factor of 2.5 for a sample of 129 galaxies in the local Universe ($z < 0.05$).
\cite{CardosoImpactAGNfeatureless2017} investigated the contribution of an AGN power-law component to spectral fitting and found that neglecting a power-law component can lead to an overestimation of $\sim 2$\,dex in stellar mass. 
They also found that stellar age was more significantly impacted by a power-law component than the stellar metallicity, but note that these biases become more severe with increasing AGN fraction.

It is therefore important to characterise the impact of AGN on the other galaxy properties as implementations of SED fitting codes often neglect the contribution of AGN (i.e. \citealt{GAMA_MAGPHYS,CarnallHowMeasureGalaxy2019,LejaHowMeasureGalaxy2019,BellstedtGalaxyMassAssembly2020b,BellstedtGalaxyMassAssembly2021,ThorneDeepExtragalacticVIsible2021}). 
If the exclusion of AGN in our SED fitting process were to significantly impact the other properties, then this would be a significant caveat to the work. 
This is not usually an issue in the low-$z$ Universe ($z<0.1$), because AGN are expected to be rare, but the number density of AGN increases by a factor of 10 by $z=1$ and by a factor of 100 by $z=2$ \citep[e.g.][]{RichardsSloanDigitalSky2006,SmolcicCosmicEvolutionRadio2009}. 
Therefore, to recover accurate galaxy properties using photometry at higher redshifts, it is essential to ensure that we are not creating any specific biases, by adequately characterising the impact of AGN on SED science. 

We use the SED fits for the same sample of galaxies from \cite{ThorneDeepExtragalacticVIsible2021} to investigate the impact of a significant AGN component on various galaxy properties. 
The fits from \cite{ThorneDeepExtragalacticVIsible2021} were run without an AGN component using the same star formation and metallicity history prescriptions, the same dust models and priors, and the same photometry. 
The only minor changes between the AGN and no AGN versions are a slightly longer chain for the fitting when including an AGN, and including photometry within the PAH features which was excluded when fitted with no AGN component (as discussed in Section~\ref{sec:AGNComp}). 
Photometry within the PAH features was removed in \cite{ThorneDeepExtragalacticVIsible2021} as these are highly susceptible to modelling assumptions, and because of potential AGN contamination in the MIR which was not included in the modelling at the time. 
Including all MIR bands where available when fitting impacts the derived SFR and dust properties, but is minimal ($<0.01\,$dex) compared to the AGN contribution.

Figure~\ref{fig:GalChanges} shows how key galaxy properties change as a function of $f_\text{AGN}$ when the AGN model is turned on. 
We only show galaxies with $f_\text{AGN} > 0.01$ in this figure, but note that most of our sample has $f_\text{AGN} \ll 0.01$ and lies off the left of the figure, and with properties that, on average, do not change with the inclusion of AGN in the fitting.
We show a histogram of the change in each parameter on the right of each panel with the whole sample shown as the black line, and the population of sources with $f_\text{AGN}>0.1$ shown in the blue line.
This shows that for the majority of objects, the stellar mass, SFR, final metallicity, and stellar age do not change significantly on average due to the relatively low prevalence of AGN.  
Figure~\ref{fig:GalChanges} also demonstrates that including an AGN component in fits does not bias derived properties for galaxies with no significant AGN component.

As expected, we find the largest differences in galaxy properties for galaxies with $f_\text{AGN}>0.1$, in which the AGN component begins to dominate the MIR emission. 
The changes between fitting with and without an AGN contribution arise because the SED model is largely unable to model an excess of emission in the MIR (see Figure \ref{fig:SED_Example}) without an AGN component. 

An increasing $f_\text{AGN}$ reduces the SFR by up to 2 dex (as $f_\text{AGN} \rightarrow 1$) as the UV-optical power-law component of the AGN model can be degenerate with emission from young stars. 
When an AGN component is included and constrained by emission in the MIR, this degeneracy can be broken. 
Without the inclusion of the AGN model, many of the strong AGN galaxies are fitted with an extremely high recent SFR to compensate. 
This differs from \cite{LejaHotDustPanchromatic2018} who find that an increasing $f_\text{AGN}$ has a strong, though variable, effect on specific SFR. 
The differences between these findings could originate from the UV-NIR power-law component expected from AGN which is not incorporated by \cite{LejaHotDustPanchromatic2018} but is included in this work. 
We find no change in the stellar age (shown here using the mass weighted age, as derived using equation 10 from \citealt{MacArthurStructureDiskdominatedGalaxies2004}) except at the very highest values of $f_\text{AGN}$.

We also find no considerable impact on the stellar mass ($<0.05$\,dex on average) and that a high $f_\text{AGN}$ (> 0.5) decreases the final gas phase metallicity by up to 0.2 dex on average. 
The dust mass is the most affected by a large $f_\text{AGN}$ as the AGN model can both reduce the attenuation required in the UV-NIR through the contribution of the power-law component and reduce the emission required in the FIR through the contribution of the dusty torus. 
As stated in \cite{ThorneDeepExtragalacticVIsible2021} the dust mass estimates have large model-dependent uncertainties that are not truly reflected in the uncertainties from the fitting process and are treated as a nuisance parameter.  
For all parameters at all values of $f_\text{AGN}$ there is some scatter around zero due to randomness in the optimisation, the longer optimisation chain and the inclusion of the MIR bands within the PAH features. 

In Figure~\ref{fig:GalChangesWithz} we also show the redshift evolution of the median offset in each of the parameters shown in Figure~\ref{fig:GalChanges} for objects with no AGN contribution ($f_\text{AGN} < 0.1$, black) and for objects with an AGN contribution ($f_\text{AGN} > 0.1$, blue). 
We find no offset in the stellar mass, metallicity or stellar age for objects with no AGN contribution at all redshifts.
We do find that SFRs are slightly higher (0.02 dex on average) for these objects for $z>0.5$. 
This is due to the inclusion of the bands covering the PAH features in the fitting when fitting with the AGN model. 
The inclusion of these bands also impacts the recovered $\tau_\text{screen}$ and dust masses. 

For objects with $f_\text{AGN} > 0.1$ we find an offset in all parameters in at least one epoch.
The stellar mass offset increases with redshift in a step-like manner because of two effects.
The first of these is that as the Lyman break (912{\AA}) moves through the UV and optical bands, there are fewer photometric measurements of the SED, resulting in a reduced constraint of the SED fit. 
The second effect is that as the AGN-dominated portion of the SED moves into the observed-frame FIR with redshift, the data are shallower also reducing the constraint. 
However, the stellar mass offset remains at less than 0.2\,dex for $z<4$.
The final gas phase metallicity (\texttt{Zfinal}) and stellar age are most impacted by the inclusion of an AGN model at low redshift ($z<0.5$) but show little change at higher redshifts ($z>1$).
We find the largest offset for the dust parameters as per Figure~\ref{fig:GalChangesWithz}, which is constant at $\sim0.2$ dex for $z > 1$.
This $\sim0.2$\,dex offset is most likely driven by the imposed dust priors, as beyond $z > 1$ there are fewer objects with measured FIR photometry which significantly reduces the constraint on an AGN component.
Most significantly, at decreasing redshifts ($z<1$) we find an increasing offset in SFR ranging from 0.2\,dex at $z=1$ to $\approx 0.9$\,dex at $z=0.2$. 
\newline

We conclude that incorporating the flexibility of an AGN component in the SED fitting of galaxies with no significant AGN contribution has no significant impact on the derived galaxy properties on average.
We stress, however, that in order to obtain accurate estimates of the properties of an AGN's host galaxy, including an AGN model is crucial especially at higher redshifts. 

\section{AGN Luminosity Function}\label{sec:AGNLF}
\begin{figure*}
    \centering
    \includegraphics[width = \linewidth]{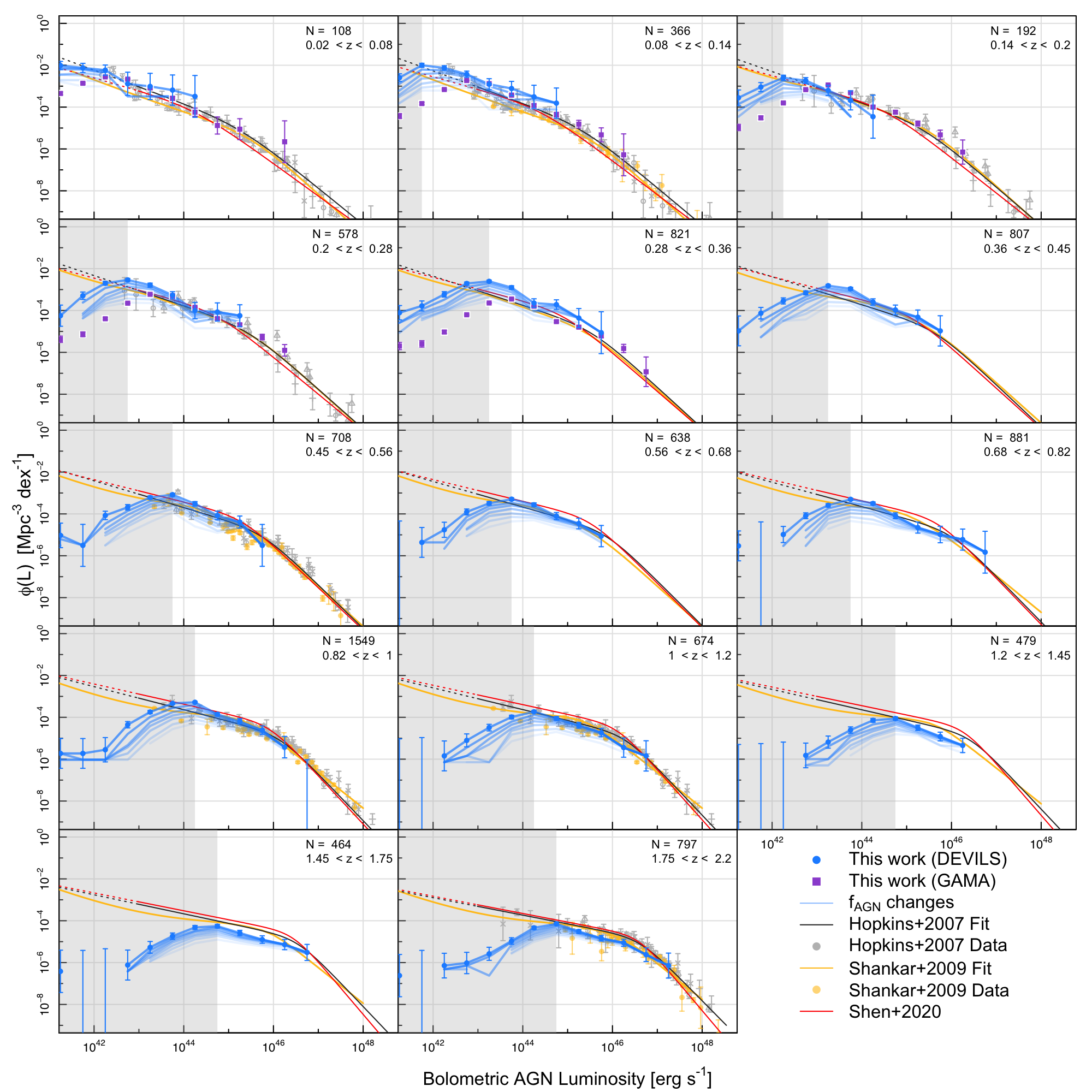}
    \caption{Bolometric AGN luminosity functions at 14 redshift bins, selected to be evenly spaced in lookback time.
    We show the luminosity function from DEVILS in the solid blue points and the low redshift GAMA results as the purple squares.
    We also show the impact of changing the required $f_\text{AGN}$ threshold as the blue lines showing steps from $f_\text{AGN} = 0.1 $ (darkest) to $0.9$ (lightest) in increments of $0.1$.
    The grey shaded region shows the range of luminosities for which we are incomplete in DEVILS.
    The number of AGN in each redshift bin is shown in the upper right, with the redshift limits of each panel. 
    We compare our results to previous bolometric luminosity functions from \citet{HopkinsObservationalDeterminationBolometric2007} (grey), \citet{ShankarSELFCONSISTENTMODELSAGN2009} (yellow), and \citet{Shenbolometricquasarluminosity2020} (red) which use a compilation of observational data in the rest-frame IR, B band, UV, soft and hard X-ray.
    }
    \label{fig:AGNLF}
\end{figure*}

\begin{figure}
    \centering
    \includegraphics[width = \linewidth]{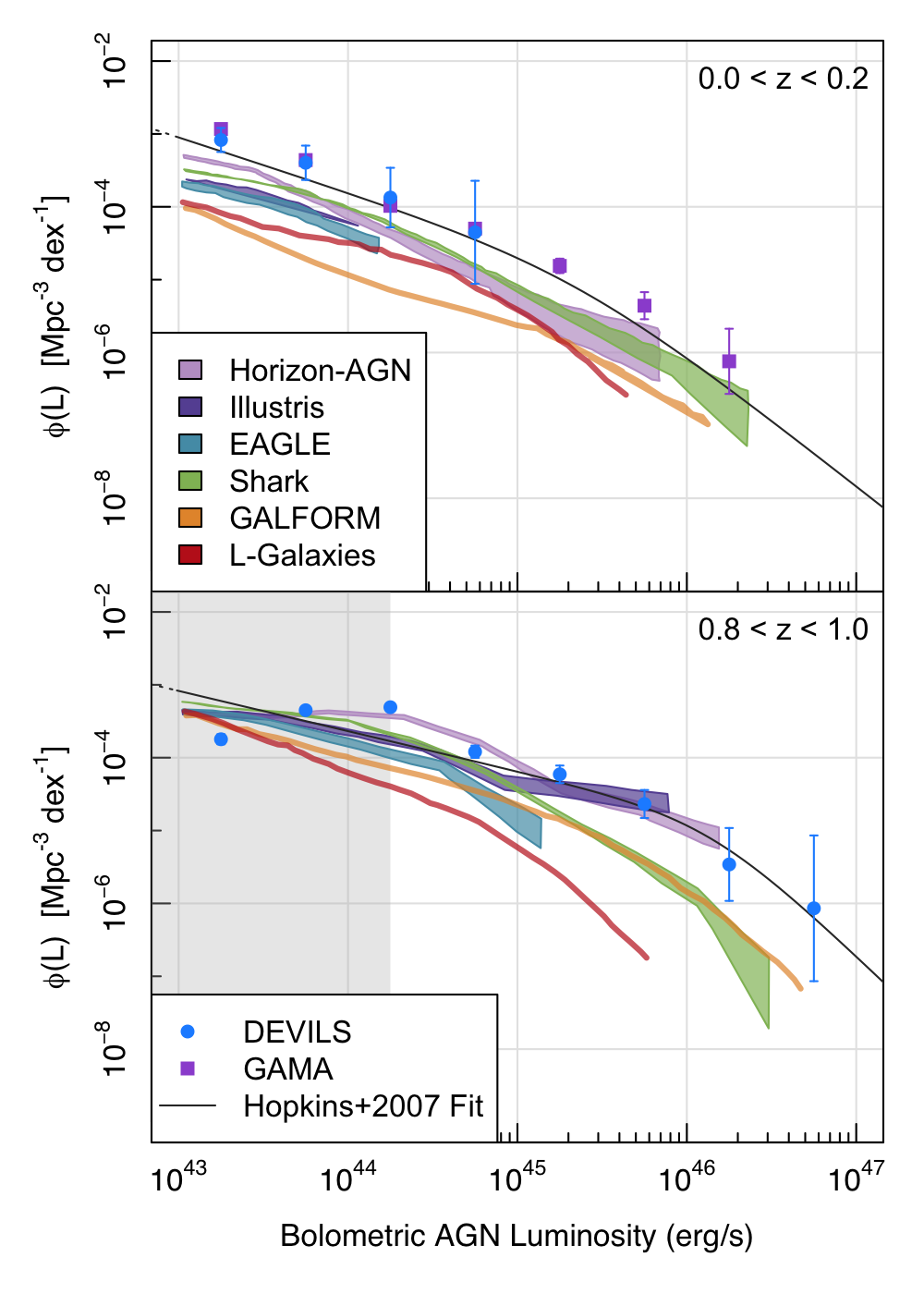}
    \caption{Bolometric AGN luminosity function at low (\textit{top panel}) and intermediate (\textit{bottom panel}) redshift compared to luminosity functions derived in \citet{Amarantidisfirstsupermassiveblack2019} from various simulations. 
    The simulation luminosity functions have been converted from X-ray luminosity functions using the bolometric corrections from \citet{HopkinsObservationalDeterminationBolometric2007}. 
    We show comparisons to Horizon-AGN \citep{DuboisDancingdarkgalactic2014} in light purple, Illustris \citep{VogelsbergerIntroducingIllustrisProject2014} in dark purple, \textsc{eagle} \citep{CrainEAGLEsimulationsgalaxy2015,SchayeEAGLEprojectsimulating2015} in blue, \textsc{Shark} \citep{LagosSharkintroducingopen2018} in green, GALFORM \citep{ColeHierarchicalgalaxyformation2000,Laceyunifiedmultiwavelengthmodel2016} in orange, and L-Galaxies \citep{HenriquesGalaxyformationPlanck2015} in red.
    We show the DEVILS (blue) results for both redshift bins and GAMA (purple) for low redshift, and shade the region in the higher redshift bin where our results become incomplete. 
    We also show the fit from \citet{HopkinsObservationalDeterminationBolometric2007} for comparison.
    }
    \label{fig:AGNLFSims}
\end{figure}

To further demonstrate the power of AGN identified and quantified with \textsc{ProSpect}, and to further validate the luminosities derived, we construct the AGN luminosity function for $0.02 < z < 2.2$.
Figure~\ref{fig:AGNLF} shows our recovered AGN luminosity function in bins evenly spaced in lookback time out to $z=2.2$ as beyond this we diverge significantly from previous measurements due to incompleteness. 
In a luminosity and redshift bin, we define the binned luminosity function as
\begin{equation}
    \phi = \frac{N_\text{AGN}}{V_\text{bin}},
\end{equation}
where $N_\text{AGN}$ is the number of AGN in a given luminosity and redshift bin, and $V_\text{bin}$ is the comoving volume of the redshift bin. 
In each redshift bin we estimate the uncertainty in the luminosity function by re-sampling the AGN luminosities from the final MCMC chain 2000 times and extracting the 16th and 84th percentiles in each luminosity bin. 
We combine this in quadrature with the Poisson uncertainties, which are the dominant source of uncertainty in most cases. 

The resultant binned luminosity function estimates are shown by the blue points in Figure~\ref{fig:AGNLF}.

When constructing the AGN luminosity function, it is possible that the number densities will be contaminated by objects with unconstrained AGN luminosities.
This is especially problematic at the high luminosity end where there are a low number of sources. 
These unconstrained AGN are caused by a lack of MIR-FIR photometry for the object due to the two-stage method of FIR photometry extraction implemented for the DEVILS catalogues and described in detail in \cite{DaviesDeepExtragalacticVIsible2021}.
Briefly, FIR photometry is only extracted for objects that have  Y$ < 21.2$ mag or are detected in the MIPS 24 imaging.
This means that for objects that do not satisfy either criteria, no attempt is made at extracting measurements in any FIR band. 
This results in no constraint on the shape of the SED in the FIR and ultimately leaves the AGN component unconstrained in many cases. 
In Figure~\ref{fig:AGNLF} we limit the sample to only sources that were passed to the FIR photometry stage and have $f_\text{AGN} >0.1$ so that any AGN component impacts the shape of the SED and has constraint provided by the photometry.
We also limit the redshift range shown in Figure~\ref{fig:AGNLF}, as beyond $z>2.2$ there are very few objects with FIR photometry due to the depth of the imaging resulting in very little constraint on an AGN component. 

When constructing the AGN luminosity function, we do not include completeness corrections but acknowledge that at low luminosities we will be incomplete for two reasons. 
Firstly, at low-luminosities it is difficult to differentiate AGN emission from that of the host galaxy and \textsc{ProSpect} will not recover an AGN component.
Secondly, we will be incomplete to low luminosity galaxies in general due to the imposed FIR requirement. 
As FIR photometry is only measured for sources with $Y<21.2$\,mag or a MIPS24 detection, the faintest galaxies (least massive or least star-forming) at each redshift will be missed, and also their potential AGN.
The impact of incompleteness on the derived luminosity function is evident in Figure~\ref{fig:AGNLF} through the turn-over in density at all redshifts. 
We show the luminosity range where we are incomplete as the grey shaded region.

We show comparisons to literature luminosity functions from \cite{HopkinsObservationalDeterminationBolometric2007,ShankarSELFCONSISTENTMODELSAGN2009, Shenbolometricquasarluminosity2020} who use compilations of optical, X-ray, and IR selected AGN to measure the bolometric quasar luminosity out to $z=6$.
We find clear evidence for evolution in the normalisation with redshift and find good agreement with literature measurements at the high luminosity end at all redshifts. 
Due to the small area covered by D10 and the lack of bright X-ray in the D10-COSMOS field at low redshift, we do not recover extremely bright AGN at low redshifts.
As described in Section~\ref{sec:NLBL_GAMA}, we have also fit the low redshift, larger area sample of galaxies in the GAMA regions using almost the same prescription as for DEVILS.
We show these results as the open circles out to $z=0.36$ and find very good agreement with both the DEVILS measurements but also the previous literature measurements. 
Using GAMA at low redshifts allows us to trace the AGN luminosity function out to AGN luminosities $\sim1\,$dex higher than DEVILS (in one case 2 dex higher). 
Supplementing the low-redshift DEVILS results with GAMA also limits the impact of cosmic variance which is high for DEVILS due to the very low area. 
We also show the impact of changing our $f_\text{AGN}$ cut as the blue lines, which we show in increments of 0.1 where the darker blues represent lower values of $f_\text{AGN}$.
At the highest luminosities this makes little difference, but at intermediate luminosities this can cause a spread of up to $\sim1\,$dex.

Finally, we show comparisons of our bolometric luminosity function to theoretical models in Figure~\ref{fig:AGNLFSims} in two redshift bins ($0.0 < z < 0.2$ and $0.8 < z < 1.0$). 
We compare to theoretical predictions from a range of hydrodynamical simulations and semi-analytic models presented in \cite{Amarantidisfirstsupermassiveblack2019}. 
To calculate the luminosity functions, \cite{Amarantidisfirstsupermassiveblack2019} estimate the bolometric luminosity for each SMBH present. 
To do this they distinguish between quasar accretion mode (thick disc scenario) and the radio accretion mode (which is assumed to take place whenever the accretion rate is below 1 per cent of the Eddington accretion limit).
The bolometric luminosities are calculated using the prescription from \cite{GriffinevolutionSMBHspin2019} accounting for these different accretion modes. 
The luminosity functions presented in \cite{Amarantidisfirstsupermassiveblack2019}, are done so as a function of X-ray luminosity.
To convert back to a bolometric luminosity, we use the \cite{HopkinsObservationalDeterminationBolometric2007} correction instead of the \cite{NetzerBolometriccorrectionfactors2019} correction used in Section~\ref{sec:xray} as this was the correction used in \cite{Amarantidisfirstsupermassiveblack2019}. 
We show values from the hydrodynamical simulations Horizon-AGN \citep{DuboisDancingdarkgalactic2014}, Illustris \citep{VogelsbergerIntroducingIllustrisProject2014}, \textsc{eagle} \citep{CrainEAGLEsimulationsgalaxy2015,SchayeEAGLEprojectsimulating2015}, and the semi-analytic models \textsc{Shark} \citep{LagosSharkintroducingopen2018}, GALFORM \citep{ColeHierarchicalgalaxyformation2000,Laceyunifiedmultiwavelengthmodel2016}, and L-Galaxies (also known as the Munich model; \citealt{HenriquesGalaxyformationPlanck2015}). 
Of the hydrodynamical models shown, Horizon-AGN is the only model that specifically tracks the spin of the SMBH. 
GALFORM uses the accretion of gas and the transfer of angular momentum to change the spin of the SMBH, which is used to calculate a spin-dependent radiative efficiency.
\citet{Amarantidisfirstsupermassiveblack2019} note that their luminosity function estimates should be considered only as lower limits, due to the limited volume of the simulations and the inability to reproduce the most extreme SMBH masses.

We find a higher number density of AGN at all luminosities at low redshift ($z < 0.2$) in both DEVILS and GAMA than predicted by the simulations, with both samples predicting a higher density than \cite{HopkinsObservationalDeterminationBolometric2007}.
We find closer agreement with the simulations at higher redshift ($0.8 < z < 1.0$) especially with Horizon-AGN. 
When presenting the theoretical X-ray luminosity functions, \cite{Amarantidisfirstsupermassiveblack2019} find good agreement with observed X-ray luminosity functions from \citet{AirdNuSTARExtragalacticSurvey2015} which predicts lower densities of AGN than DEVILS or GAMA. 
This could be due to an underestimation of Compton-thick sources predicted by \cite{AirdNuSTARExtragalacticSurvey2015} who assume a Compton-thick fraction of 25 per cent, while \cite{UedaStandardPopulationSynthesis2014} assume a fraction of 50 per cent. 
Overall, the agreement in the luminosity function constructed from \textsc{ProSpect}-derived AGN luminosities with those from observations and theoretical predictions suggests that we can recover a significant fraction of high luminosity AGN using SED fitting. 
We are unable to recover the luminosity function at low luminosities due to the inability to distinguish a low luminosity AGN component from its host galaxy and due to incompleteness in the general galaxy sample.

\section{Conclusions}\label{sec:conclusion}
We have applied the \textsc{ProSpect} SED-fitting code with an incorporated AGN component to 494,000 galaxies from the D10 field of the DEVILS survey and 230,000 galaxies from the GAMA survey to identify and quantify AGN. 
We combine a parametric star formation history and evolving metallicity tied to the growth of stellar mass with an AGN component to extract AGN and host galaxy properties for each galaxy. 
The results are summarised as follows:
\begin{itemize}
    \item In this work we obtain AGN luminosities for 9761 galaxies in the DEVILS D10 field with a fractional MIR flux contribution from the AGN component of $f_\text{AGN} > 0.1$ and with FIR constraint. 
    \item We also recover an AGN component in 67,258 galaxies in the GAMA G09, G12, G15, and G23 fields.
    \item We demonstrate that while these \textsc{ProSpect}-selected AGN agree well with MIR colour-colour selections, the SED selected sample is more reliable and far less likely to be contaminated by galaxies with no AGN component.
    \item We also show that our derived luminosities are consistent with X-ray measurements from \textit{Chandra} and previous SED fits to X-ray selected AGN. 
    \item We find that \textsc{ProSpect} identifies a significant AGN component ($f_\text{AGN} > 0.1$) in 91 per cent of AGN selected through the presence of narrow and broad emission lines. 
    \item On average, host galaxy properties such as stellar mass, stellar age, and final gas phase metallicity do not change significantly between SED fits with and without an included AGN component. However, for objects with a significant MIR AGN contribution, we find that the SFR can be up to 2\,dex lower when we allow for an AGN component in the SED fitting.
    \item Using the AGN luminosities recovered from \textsc{ProSpect} for both the DEVILS and GAMA, samples we find good agreement with previous bolometric luminosity functions from \cite{HopkinsObservationalDeterminationBolometric2007,ShankarSELFCONSISTENTMODELSAGN2009,Shenbolometricquasarluminosity2020} for $ 0.02 < z < 2$. 
    Due to the lack of very bright AGN ($L_\text{AGN} > 10^{47}\,\text{erg s}^{-1}$) in DEVILS and GAMA, we are unable to constrain the luminosity density of AGN without much larger and deeper surveys. 
    Given the increased area and depth of the upcoming Wide Area Vista Extragalactic Survey (WAVES; \citealt{Driver4MOSTConsortiumSurvey2019}), it could be possible to constrain the evolution with redshift of the AGN luminosity density using SED fitting alone.
    \item We find that our \textsc{ProSpect} derived luminosity functions are systematically higher than predictions from simulations at $z\sim0.1$ and $z\sim0.9$.
\end{itemize}
From this we conclude that SED fitting using \textsc{ProSpect} is a viable method of AGN identification and can recover AGN luminosities consistent with other methods. 
We also conclude that in order to obtain accurate galaxy properties for sources with a significant AGN component, including an AGN model in the SED fitting is imperative. 

The radio emission and host galaxies properties of these SED-selected AGN will be explored in detail in Thorne et al. (in prep).

\section{Data availability}
The DEVILS and GAMA data products described in this paper are currently available for internal team use for proprietary science and will be made available in upcoming data releases.
The SED fits to the \cite{Brownspectralenergydistributions2019} sample will be shared upon reasonable request to the corresponding author. 

\section*{Acknowledgements}
We thank the referee for their comments which have significantly improved the quality of this work. 
JET is supported by the Australian Government Research Training Program (RTP) Scholarship.
ASGR and LJMD acknowledge support from the \textit{Australian Research Council's} Future Fellowship scheme (FT200100375 and FT200100055 respectively). 
SB and SPD acknowledge support from the \textit{Australian Research Council's} Discovery Project scheme (DP180103740). 
DEVILS is an Australian project based around a spectroscopic campaign using the Anglo-Australian Telescope. The DEVILS input catalogue is generated from data taken as part of the ESO VISTA-VIDEO \citep{JarvisVISTADeepExtragalactic2013} and UltraVISTA \citep{McCrackenUltraVISTAnewultradeep2012} surveys. DEVILS is part funded via Discovery Programs by the Australian Research Council and the participating institutions. The DEVILS website is \url{https://devilsurvey.org}. The DEVILS data is hosted and provided by AAO Data Central (\url{https://datacentral.org.au/}).

GAMA is a joint European-Australasian project based around a spectroscopic campaign using the Anglo- Australian Telescope. The GAMA input catalogue is based on data taken from the Sloan Digital Sky Survey and the UKIRT Infrared Deep Sky Survey. Complementary imaging of the GAMA regions is being obtained by a number of in-dependent survey programmes including GALEX MIS, VST KiDS, VISTA VIKING, WISE, Herschel-ATLAS, GMRT and ASKAP providing UV to radio coverage. GAMA is funded by the STFC (UK), the ARC (Australia), the AAO, and the participating institutions. The GAMA website is \url{http://www.gama-survey.org/}.

This work was supported by resources provided by the Pawsey Supercomputing Centre with funding from the Australian Government and the Government of Western Australia.
We gratefully acknowledge DUG Technology for their support and HPC services.

All of the work presented here was made possible by the free and open R software environment \citep{RCoreTeamLanguageEnvironmentStatistical2020}. All figures in this paper were made using the R \textsc{magicaxis} package \citep{RobothammagicaxisPrettyscientific2016}. This work also makes use of the \textsc{celestial} package \citep{RobothamCelestialCommonastronomical2016}. 




\bibliographystyle{mnras}
\bibliography{MyBib} 



\appendix

\section{Fits to the Brown et al. (2019) Atlas of AGN SEDs }\label{app:Brown}

\begin{table*}
    \centering
    \caption{Values of AGN parameters for fits to the sample of AGN presented in \citet{Brownspectralenergydistributions2019}. Note the 	\texttt{AGNlum}	parameter is $\log{10} (L_\text{AGN} / \text{erg s}^{-1})$, the \texttt{AGNta} is logged as per fitting, and  \texttt{AGNan}$=0\deg$ when viewed edge on through the torus and \texttt{AGNan}$=90\deg$ when viewed face-on.}
    \label{tab:BrownAtlas}
    \small
    \begin{tabular}{l l c c c c}
    \hline
        Name	&	z	&		\texttt{AGNlum}			&		\texttt{AGNta}			&		\texttt{AGNan}				&	Reduced $\chi^2$		\\
        \hline
2MASX J13000533+1632151	&	0.0799	&	$	45.82	\pm	0.05	$	&	$	0.76	\pm	0.11	$	&	$	32	\pm	5	$	&	1.054		\\
3C 120	&	0.033	&	$	45.33	\pm	0.03	$	&	$	0.675	\pm	0.16	$	&	$	43	\pm	2	$	&	4.891		\\
3C 273	&	0.1583	&	$	46.96	\pm	0.04	$	&	$	0.659	\pm	0.22	$	&	$	43	\pm	2	$	&	4.481		\\
3C 351	&	0.3719	&	$	46.89	\pm	0.05	$	&	$	0.762	\pm	0.22	$	&	$	46	\pm	1	$	&	3.669		\\
3C 390	&	0.0561	&	$	45.31	\pm	0.04	$	&	$	0.756	\pm	0.19	$	&	$	46	\pm	1	$	&	5.677		\\
Ark 120	&	0.0327	&	$	45.38	\pm	0.02	$	&	$	0.384	\pm	0.12	$	&	$	44	\pm	1	$	&	1.544		\\
Ark 564	&	0.0247	&	$	44.75	\pm	0.06	$	&	$	0.911	\pm	0.37	$	&	$	38	\pm	10	$	&	1.312		\\
F2M1113+1244	&	0.6812	&	$	47.67	\pm	0.05	$	&	$	0.488	\pm	0.19	$	&	$	28	\pm	6	$	&	1.789		\\
Fairall 9	&	0.047	&	$	45.66	\pm	0.03	$	&	$	0.782	\pm	0.15	$	&	$	44	\pm	1	$	&	2.869		\\
H 1821+643	&	0.2968	&	$	47.19	\pm	0.04	$	&	$	0.67	\pm	0.25	$	&	$	40	\pm	6	$	&	5.596		\\
IRAS 11119+3257	&	0.1876	&	$	46.8	\pm	0.03	$	&	$	0.489	\pm	0.17	$	&	$	33	\pm	4	$	&	0.414		\\
IRAS F16156+0146	&	0.132	&	$	44.9	\pm	0.15	$	&	$	0.783	\pm	0.74	$	&	$	41	\pm	1	$	&	5.357		\\
Mrk 110	&	0.0353	&	$	44.78	\pm	0.03	$	&	$	0.679	\pm	0.14	$	&	$	44	\pm	1	$	&	1.451		\\
Mrk 1502	&	0.0589	&	$	45.9	\pm	0.03	$	&	$	0.774	\pm	0.16	$	&	$	43	\pm	1	$	&	3.580		\\
Mrk 231	&	0.0422	&	$	46.38	\pm	0.03	$	&	$	0.614	\pm	0.2	$	&	$	40	\pm	2	$	&	0.808		\\
Mrk 279	&	0.0305	&	$	44.87	\pm	0.04	$	&	$	0.856	\pm	0.16	$	&	$	46	\pm	2	$	&	2.623		\\
Mrk 290	&	0.0302	&	$	44.57	\pm	0.03	$	&	$	0.764	\pm	0.24	$	&	$	47	\pm	2	$	&	1.253		\\
Mrk 421	&	0.03	&	$	44.85	\pm	0.03	$	&	$	0	\pm	0.37	$	&	$	51	\pm	5	$	&	0.922		\\
Mrk 493	&	0.031	&	$	44.65	\pm	0.03	$	&	$	0.317	\pm	0.46	$	&	$	47	\pm	2	$	&	1.863		\\
Mrk 509	&	0.344	&	$	47.51	\pm	0.04	$	&	$	0.397	\pm	0.27	$	&	$	44	\pm	1	$	&	3.481		\\
Mrk 590	&	0.0261	&	$	44.03	\pm	0.1	$	&	$	0.467	\pm	0.72	$	&	$	40	\pm	4	$	&	4.147		\\
Mrk 817	&	0.0315	&	$	45.11	\pm	0.03	$	&	$	0.693	\pm	0.19	$	&	$	47	\pm	1	$	&	3.239		\\
Mrk 876	&	0.129	&	$	46.1	\pm	0.03	$	&	$	0.777	\pm	0.18	$	&	$	44	\pm	1	$	&	2.682		\\
Mrk 926	&	0.0469	&	$	45.44	\pm	0.03	$	&	$	0.301	\pm	0.14	$	&	$	41	\pm	1	$	&	2.227		\\
NGC 3227 Central	&	0.0039	&	$	43.34	\pm	0.12	$	&	$	0.972	\pm	0.43	$	&	$	33	\pm	5	$	&	3.188		\\
NGC 3516 Central	&	0.0088	&	$	44.09	\pm	0.04	$	&	$	1	\pm	0.12	$	&	$	45	\pm	2	$	&	3.099		\\
NGC 4051 Central	&	0.0023	&	$	43.02	\pm	0.04	$	&	$	-0.753	\pm	0.25	$	&	$	41	\pm	4	$	&	0.969		\\
NGC 4151 Central	&	0.0033	&	$	43.83	\pm	0.03	$	&	$	0.934	\pm	0.11	$	&	$	45	\pm	1	$	&	3.515		\\
NGC 5548 Central	&	0.0166	&	$	44.35	\pm	0.03	$	&	$	0.549	\pm	0.47	$	&	$	43	\pm	1	$	&	3.432		\\
NGC 5728	&	0.0094	&	$	37.6	\pm	3.55	$	&	$	0.819	\pm	1	$	&	$	52	\pm	7	$	&	0.836		\\
NGC 7469	&	0.0163	&	$	44.76	\pm	0.03	$	&	$	0.301	\pm	0.4	$	&	$	49	\pm	3	$	&	4.767		\\
OQ 530	&	0.1525	&	$	46.03	\pm	0.03	$	&	$	-0.246	\pm	0.2	$	&	$	43	\pm	3	$	&	0.645		\\
PG 0026+129	&	0.142	&	$	45.8	\pm	0.04	$	&	$	0.637	\pm	0.3	$	&	$	44	\pm	3	$	&	2.374		\\
PG 0052+251	&	0.1545	&	$	45.87	\pm	0.04	$	&	$	0.778	\pm	0.19	$	&	$	46	\pm	2	$	&	3.386		\\
PG 1211+143	&	0.0809	&	$	45.77	\pm	0.06	$	&	$	1	\pm	0.04	$	&	$	42	\pm	1	$	&	5.323		\\
PG 1307+085	&	0.1538	&	$	45.78	\pm	0.03	$	&	$	-1	\pm	0.14	$	&	$	59	\pm	8	$	&	5.337		\\
PG 1415+451	&	0.1137	&	$	45.55	\pm	0.02	$	&	$	-0.241	\pm	0.34	$	&	$	46	\pm	2	$	&	1.139		\\
PG 2349-014	&	0.1738	&	$	46.08	\pm	0.03	$	&	$	0.548	\pm	0.19	$	&	$	43	\pm	1	$	&	3.691		\\
PKS 1345+12	&	0.1205	&	$	45.79	\pm	0.1	$	&	$	-0.995	\pm	0.16	$	&	$	21	\pm	4	$	&	2.090		\\
Ton 951	&	0.064	&	$	45.08	\pm	0.04	$	&	$	1	\pm	0.16	$	&	$	89	\pm	3	$	&	3.342		\\
W Com	&	0.102	&	$	45.76	\pm	0.02	$	&	$	-0.598	\pm	0.24	$	&	$	38	\pm	3	$	&	0.721		\\
\hline
    \end{tabular}
\end{table*}

We also fit the atlas of AGN SEDs presented in \cite{Brownspectralenergydistributions2019} using the same procedure presented in Section~\ref{sec:SED}.
The \cite{Brownspectralenergydistributions2019} atlas includes the SEDs of 41 low-redshift well studied AGN with coverage of at least 0.09 to 30\,$\mu$m but in some cases the SEDs extend into the FIR.
As the photometry for each of the sources are obtained from various facilities with differing filter sets we elect to use the synthesised SEDs which were created by filling the gaps in spectral coverage to map the SED onto the same filter set. 
For all galaxies we use the synthesised photometry in the GALEX FUV NUV, SDSS ugriz, 2MASS JHKs, IRAC 1-4, WISE 1-4, MIPS 24 70 bands and the PACS 100, 160 and SPIRE 250,350 bands where available.
As no photometric errors are provided we assume a 0.05\,mag error in the FUV-W4 bands and a 0.1 mag error in the FIR to mimic the uncertainties measured from the DEVILS photometry. 
We acknowledge that these sources have significantly more photometric coverage across the MIR than the DEVILS sources due to the inclusion of both Spitzer IRAC and WISE measurements. 
Due to the way the sample was selected, the sample favours bright quasars or nearby galaxy nuclei that are dominated by AGN light. 
As the AGN component can outshine the contribution from the host galaxy (i.e. $f_\text{AGN} > 0.9$), there is very little constraint on the host galaxy properties in the SED and we therefore recover unphysical solutions for the stellar mass and star formation of the host galaxy. The right panel of Figure~\ref{fig:3C273} shows the fit to Ark 120 using the implementation described above.

Table~\ref{tab:BrownAtlas} includes the recovered AGN properties for each of these sources and the reduced $\chi^2$ of the SED fit. 
Note that we have not explored the accuracy of the \texttt{AGNta} or \texttt{AGNan} parameters. 
In the implementation of the \cite{FritzRevisitinginfraredspectra2006} and \cite{FeltreSmoothclumpydust2012} model in \text{ProSpect} the shape of the SED for the AGN component changes for angles between 0-30\,degrees due to increased attenuation through the torus, but above 30 degrees there is no difference in SED because of the adopted opening angle (\texttt{AGNct} = 100 degrees).


\bsp	
\label{lastpage}
\end{document}